\documentclass[12pt, a4paper]{article}
\pdfoutput=1
\usepackage[utf8]{inputenc}
\usepackage{fancybox}
\usepackage[T1]{fontenc}
\usepackage{lmodern, textcomp}
\usepackage{bm}
\usepackage[american]{babel}
\usepackage{csquotes}

\usepackage{eurosym}
\usepackage[nottoc]{tocbibind}
\usepackage{authblk}
\usepackage{fancyhdr}
\usepackage{xspace}

\usepackage{graphicx}
\usepackage{subcaption}
\usepackage{float}
\usepackage{makecell}
\usepackage{multirow}
\usepackage{multicol} 
\usepackage{marginnote}
\usepackage[multiple]{footmisc}

\usepackage[usenames, dvipsnames, svgnames, table]{xcolor}

\usepackage[backend=bibtex8, citestyle=numeric-comp, maxbibnames=99,
			sorting=none, sortcase=false, sortcites=true, giveninits=true]{biblatex}

\usepackage{amsmath, amsfonts, amssymb}
\usepackage{mathtools}
\usepackage{mathrsfs}
\usepackage{cancel}
\usepackage{braket}
\usepackage{tensor}
\usepackage{slashed}
\usepackage{siunitx}
\usepackage{listings}

\usepackage{stmaryrd}
\usepackage{caption}
\usepackage{relsize,exscale}
\usepackage{array}
\usepackage[toc,page]{appendix}

\usepackage{enumerate}

\DeclareSymbolFont{largesymbols}{OMX}{cmex}{m}{n}

\setcellgapes{1pt}
\makegapedcells
\newcolumntype{R}[1]{>{\raggedleft\arraybackslash }b{#1}}
\newcolumntype{L}[1]{>{\raggedright\arraybackslash }b{#1}}
\newcolumntype{C}[1]{>{\centering\arraybackslash }b{#1}}

\newcommand{\cI}{{\mathcal I}}

\newcommand{\J}{\mathrm{J}}

\newcommand{\beq}{\begin{equation}}
\newcommand{\eeq}{\end{equation}}
\newcommand{\bea}{\begin{eqnarray}}
\newcommand{\eea}{\end{eqnarray}}
\definecolor{mygray}{gray}{0.3}

\newcommand{\bes}{\begin{eqnarray}}
\newcommand{\ees}{\end{eqnarray}}

\newcommand\restr[2]{{
  \left.\kern-\nulldelimiterspace 
  #1 
  \vphantom{\big|} 
  \right|_{#2} 
  }}

\newcommand{\U}{\mathrm{U}}
\usepackage{multicol}
\usepackage{amssymb}

\usepackage[heightrounded, top=3.5cm, bottom=3.5cm, left=2cm, right=2cm, headheight=14pt]{geometry}

\usepackage[pdftex, breaklinks=true, linktocpage,
	colorlinks=true, urlcolor=blue, linkcolor=blue, citecolor=red
]{hyperref}

\newcommand{\email}[1]{\href{mailto:#1}{\nolinkurl{#1}}}
\newcommand{\emailfoot}[1]{\thanks{\email{#1}}}

\usepackage{marginnote}
\newcounter{draftcommentcnt}
\NewDocumentCommand{\draftcomment}{s O{red} m}{%
	\def\margnote{\IfBooleanTF{#1}{\marginnote}{\marginpar}}%
	\stepcounter{draftcommentcnt}%
	\textcolor{#2}{#3}%
	\margnote{\textcolor{#2}{$\Leftarrow$ \arabic{draftcommentcnt}}}%
}

\fancypagestyle{preprint}{
	\fancyhf{}

	\fancyhead[R]{\textsf{\preprint}}
}

\numberwithin{equation}{section}

\hypersetup{
	pdftitle={Anomalous Ward operators composition law in group field theories without closure constraint},
}

\title{Anomalous higher order Ward identities  in tensorial group field theories without closure constraint}
\author[2]{Bio Wahabou Kpera\emailfoot{wahaboukpera@gmail.com}}
\author[1]{Vincent Lahoche\emailfoot{vincent.lahoche@cea.fr}}
\author[1,2]{Dine Ousmane Samary\emailfoot{dine.ousmanesamary@cipma.uac.bj}}
\author[2]{Seke Fawaaz Zime Yerima\emailfoot{szimeyerima@gmail.com}}

\affil[1]{%
	Université Paris Saclay, \textsc{Cea}, \textsc{List}, Gif-sur-Yvette, F-91191, France
}

\affil[2]{%
	Faculté des Sciences et Techniques (ICMPA-UNESCO Chair)
	\protect\\
	Université d'Abomey-Calavi, 072 BP 50, Bénin
}

\begin{document}
\maketitle

\begin{abstract}
The Ward-Takahashi identities are considered as the generalization of the Noether currents available to quantum field theory and include quantum fluctuation effects. Usually, they take the form of relations between correlation functions, which ultimately correspond to the relation between coupling constants of the theory. For this reason, they play a central role in the construction of renormalized theory, providing strong relations between counter-terms. Since last years, they have been intensively considered in the construction of approximate solutions for nonperturbative renormalization group of tensorial group field theories. The construction of these identities is based on the formal invariance of the partition function under a unitary transformation, and Ward's identities result from a first-order expansion around the identity. Due to the group structure of the transformation under consideration, it is expected that a first-order expansion is indeed sufficient. We show in this article that this does not seem to be the case for a complex tensor theory model, with a kinetic term involving a Laplacian. 
\end{abstract}

\noindent Pacs numbers:

\setcounter{footnote}{0}
\newpage

\hrule
\pdfbookmark[1]{\contentsname}{toc}
\tableofcontents
\bigskip
\hrule

\newpage

\section{Introduction}

For nearly twenty years, group field theories (GFT) have emerged as a promising avenue for the quantization of gravitation, offering a clever solution to the Hamiltonian constraint issue encountered in loop quantum gravity (LQG). The inspirations behind GFTs are numerous \cite{Freidel_2005,baratin2012ten,https://doi.org/10.48550/arxiv.1110.5606,https://doi.org/10.48550/arxiv.gr-qc/0607032,https://doi.org/10.48550/arxiv.1210.6257}. They can be seen as a second quantization of the LQG, itself being a canonical quantization procedure of general relativity \cite{rovelli2004quantum,https://doi.org/10.48550/arxiv.1310.7786,oriti2015group}; and the quanta of the GFTs are as many grains of space of which basic states of the LQG are collective excitations. GFTs can also be seen as effective field theories whose probability amplitudes correspond to the "spin foams" of covariant approaches to LQG \cite{Perez_2013}. Finally, from the point of view of random geometry, GFTs can be seen as attempts to generalize the random matrix approach \cite{Francesco_1995,https://doi.org/10.48550/arxiv.1510.04430} to quantum gravity in dimension 2 to all dimensions. Since the last ten years, GFTs have also benefited from the positive influence of other random geometry approaches, notably that of colored random tensors \cite{Gurau_2016,rivasseau2016random,guruau2017random}. These random tensors were initially proposed by Gurau and his collaborators as a direct extension of random matrices in dimension $d>2$, and in particular having the property of possessing a well-defined, although non-topological, $1/N$-expansion. From this point of view, tensors inspired the GFTs in the construction of their interactions, which from then on presented the same kind of tensor invariance as the random tensors. These theories are called tensorial group field theories (TGFTs) and present a well-defined power counting \cite{Carrozza_2016ccc,Carrozza_2015aaa,Carrozza_2014,rivasseau2011towards}. The shape of the quantum corrections quickly motivated the introduction of a new kind of theory, confused with TGFTs in the literature, and for which a Laplacian propagator is added to the kinetic term of the field theory. The presence of this Laplacian and the existence of a rigorous power counting allowed the implementation of a renormalization program and several BPHZ theorems have been proved \cite{samary2014just,carrozza2014renormalization2,carrozza2014renormalization,lahoche2015renormalization1}. A constructive theory program has also been undertaken, and the summability in the sense of Borel of perturbative series has been proved for some models \cite{lahoche2018constructive,lahoche2018constructive2,rivasseau2019constructive,rivasseau2016constructive,rivasseau2021can,delepouve2016constructive}. TGFTs have also allowed the construction of quantum cosmology and isolated event horizon models under the assumption of the existence of condensed states, analogous to Bose-Einstein condensates in condensed matter physics \cite{pithis2019group,jercher2022emergent,https://doi.org/10.48550/arxiv.2112.02585,de2017dynamics,kegeles2018inequivalent,oriti2017universe,oriti2016horizon}. The question of the existence of such phases for GFTs is however still an open question today. Despite more traditional approaches recently considered in \cite{pithis2018phase,marchetti2021phase} and based on a Ginsburg-Landau type analysis, the main approach considered in the literature to study phase transitions in GFTs is the nonperturbative renormalization group based on the Wetterich-Morris framework \cite{Dupuis_2021,Berges_2002}. This program started a few years ago with a series of works (see for instance \cite{Carrozza_2017,Carrozza_2017a,Lahoche_2017bb,Benedetti_2016} and reference therein), which all sets itself the goal to show a tendency to universality in the appearance of fixed points of Wilson-Fisher type. These fixed points correspond physically to non-trivial resummations of spin foams and played an important role in the existence of these hypothetical condensates. Since 2018, a program aiming at elaborating more complete methods than those considered so far, exploiting, in particular, the structure of Feynman graphs and the Ward identities inherent to the global unitary invariance of classical actions, showed that this universality is most probably a consequence of the incompleteness of the methods considered so far. In particular, the quartic and just-renormalizable models were found, in the melonic sector and beyond, to have no Wilson-Fisher fixed point \cite{Lahoche_2021c,Lahoche_2019a,Lahoche_2019bb,Lahoche_2020b,Lahoche:2018oeo}. Other works are also worth mentioning, using other methods such as effective potentials \cite{pithis2021no,pithis2020phase}. Finally, it has recently been shown that models including degrees of freedom interpretable as material particles, in a stochastic regime, do present a UV attractor compatible with a second-order phase transition \cite{lahoche2022stochastic}. On the hand of random tensors, a nonperturbative renormalization group has been proposed \cite{Eichhorn_2013,Eichhorn_2014,Eichhorn_2019,Eichhorn_2020}, in continuation of the seminal work of Brezin and Zinn-Justin \cite{Br_zin_1992}. Unfortunately, incompatibilities with Ward identities and approximations used to parametrize the theory space make the reliability of results coming from this approach an open issue \cite{Lahoche:2020pjo,lahoche2020revisited}. \\

All these results have largely demonstrated the relevance of Ward identities in studying the properties of TGFTs, particularly in that of their renormalization group. This is not a surprise, since these same identities play an important role in the renormalization of theories such as quantum electrodynamics or gauge theories, in general, \cite{peskin2018introduction,Zinn-Justin:1989rgp,ZinnJustinBook2}. So far, for TGFTs, only Ward identities resulting from a first-order variation of the internal transformation group have been considered \cite{samary2014closed2,perez2018full,pascalie2021correlation,Lahoche:2018ggd,Lahoche_2019bb}, but this was not limited in principle. Indeed, due to the group structure of the transformation under consideration, it is expected that a first-order development is sufficient because in principle any finite transformation can be deduced from the composition of infinitesimal transformations of the first-order. Thus, it was always assumed that the identities obtained in the second order or beyond should be redundant with respect to the identities in the first order because of the internal composition law of the symmetry group. In the present work, we show that this is not the case. Second-order transformations cannot be deduced from the composition of first-order transformations. While one might think that the origin of this phenomenon lies in the specific non-locality of tensor interactions, a detailed analysis shows that this is not the case. Moreover, the redundancy of Ward identities has been demonstrated for a tensor theory including a closure constraint in \cite{Lahoche:2021nba}. It seems that the origin of this phenomenon lies in the kinetic term, and in the presence or absence of constraints on the field. Indeed, it seems that this anomaly is a specificity of TGFTs without closure constraints, and we provide more details on the perspectives of this observation in the conclusion. \\

\textbf{Outline.} The paper is organized as follows. In section \ref{firstsection}, we define the model and recall the basic derivation of first-order Ward identities. In the section \ref{secondsection} section, we consider a variation of the transformation group at second order around identity and derive the second-order Ward identities. Next, we show explicitly the failure of the expected redundancy with first-order Ward identities. Finally, we discuss open issues, especially regarding interpretations in the last section \ref{thirdsection}.

\section{Preliminaries}\label{firstsection}
In this section, we provide some technical preliminaries, including a definition of the TGFT class that we consider, a derivation of the first-order Ward identities and a short review of the composition of Ward operators in standard quantum field theory. 

\subsection{Definition of the model}

\begin{figure}
\begin{center}
\includegraphics[scale=1]{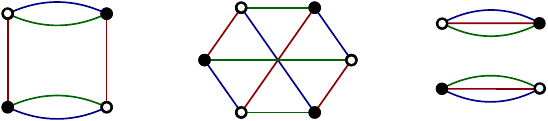}
\end{center}
\caption{Example of tensorial invariants for $d=3$}\label{figBubbles}
\end{figure}

Let us consider a Lie group $\mathbf{G}$. A group field $\varphi$ is a mapping acting on $d$ copies of the group $\mathbf{G}$:
\begin{equation}
\varphi:(\mathbf{G})^{\times d}\to \mathbb{R}\,, \mathbb{C}\,. 
\end{equation}
We denote as $d$ the rank of the group field and we call the structure group the group $\mathbf{G}$. In this paper, we consider a complex field with group structure $\mathbf{G}=\U(1)$. A GFT is a statistical field theory (SFT) for such a group field, which differs from ordinary SFT by the specific non-locality of the interactions. For TGFTs, interactions are assumed to be invariant concerning unitary transformations:
\begin{equation}
\varphi(g_1,\cdots,g_d)=\int d\mathbf{g}^\prime\, \mathcal{U}_{\textbf{c}}(\mathbf{g},\mathbf{g}^\prime) \varphi(g_1^\prime,\cdots,g_d^\prime)=: \mathcal{U}_{\textbf{c}}\triangleright
\varphi(g_1,\cdots,g_d) \,,\label{unitary}
\end{equation}
where $\textbf{c}\subset \llbracket 1,d \rrbracket$, $\bm g:= (g_1,\cdots,g_d)$, $d\mathbf{g}:=\prod_{\ell=1}^d dg_\ell$ is the standard Haar measure over $\mathbf{G}$ and:
\begin{equation}
\mathcal{U}_{\textbf{c}}(\mathbf{g},\mathbf{g}^\prime):= \prod_{c\in {\textbf{c}}}U_c(g_c,g_c^\prime) \prod_{\ell \notin {\textbf{c}}} \delta(g_\ell^\prime g^{-1}_\ell)\,,\label{unitarygen}
\end{equation}
the matrix $U_c(g_c,g_c^\prime)$ being unitary:
\begin{equation}
\int dh_c\, U_c^\dagger(g_c,h_c) U_c(h_c,g_c^\prime)=\delta(g_c^\prime g_c^{-1})\,,
\end{equation}
and we denote by $\mathbb{U}(\mathbf{G})$ the group of unitary matrices over $\mathbf{G}$. Ordinary, a GFT looks like an equilibrium SFT, and the probability density $\rho[\varphi,\bar{\varphi}]$ for a given configuration $(\varphi,\bar{\varphi})$ of fields takes the form of a Boltzmann weight:
\begin{equation}
\rho[\varphi,\bar{\varphi}] \sim e^{-S[\varphi,\bar{\varphi}]}\,,
\end{equation}
and the correlations of the GFT are deduced from the generating functional $Z[J,\bar{J}]$:
\begin{equation}
Z[J,\bar{J}]=\int [d\varphi] [d\bar{\varphi}] \, \rho[\varphi,\bar{\varphi}] \exp \left[\int d\bm g  \bar{J}(\bm g) \varphi(\bm g)+\int d\bm g \bar{\varphi}(\bm g) J(\bm g) \right]\,,\label{partitionfunction}
\end{equation}
where $[d\varphi] [d\bar{\varphi}]$ is the path integral measure over the structure group manifold.  For the rest of this paper it will be convenient to introduce the dot product:
\begin{equation}
A \cdot B = \int d\bm g\, A(\bm g) B(\bm g)\,.
\end{equation}
The \textit{classical action} $S[\varphi,\bar{\varphi}]$ is assumed to be of the generic form:
\begin{equation}
S[\varphi,\bar{\varphi}]= \int d\bm g d\bm g^\prime\, \bar{\varphi}(\bm g) \mathcal{K}(\bm g, \bm g^\prime) \varphi(\bm g^\prime)+S_{\text{int}}[\varphi,\bar{\varphi}]\,,
\end{equation}
where for TGFTs the kinetic kernel $\mathcal{K}(\bm g, \bm g^\prime)$ is a diagonal operator involving the Laplacian over the structure group:
\begin{equation}
\mathcal{K}(\bm g, \bm g^\prime)=\delta(\bm g^\prime \bm g^{-1}) (\Delta_{\bm g}+m^2)\,,
\end{equation}
where $\Delta_{\bm g}:=\sum_c \Delta_{g_c}$, $\Delta_{g_c}$ being the Laplacian on the $c$-th copies of $\mathbf{G}$. The interaction part $S_{\text{int}}[\varphi,\bar{\varphi}]$ includes all contributions that are not Gaussian. For TGFTs, interaction is a polynomial in fields $\varphi$ and $\bar{\varphi}$, such that each monomial involves an even number of fields. Furthermore, group labels of fields are contracted such that each monomial is invariant under unitary transformations like \eqref{unitary}. One call \textit{tensorial invariant} such a monimial. These invariants admit an elegant representation in terms of $d$-colored bipartite regular graphs. The receipt to construct them is the following:
\begin{enumerate}
\item To each field $\varphi$ and $\bar{\varphi}$ we assign a black and white dot respectively, with $d$ half-colored edges hooked to them, materializing the $d$ group variables $g_1,\cdots, g_d$:
\begin{equation*}
\vcenter{\hbox{\includegraphics[scale=0.8]{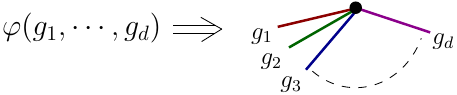}}}\qquad \vcenter{\hbox{\includegraphics[scale=0.8]{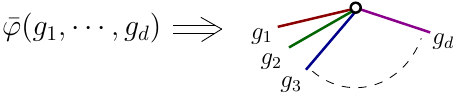}}}\,.
\end{equation*}
\item Colored edges are then hooked together, accordingly with their respective colors, between black and white dots only.
\end{enumerate}
In Figure \ref{figBubbles} we show some examples for $d=3$. To provide an explicit example, the first diagram reads explicitly as:
\begin{equation}
\vcenter{\hbox{\includegraphics[scale=0.75]{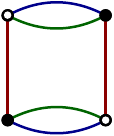}}}\equiv \int \prod_{i=1}^3 dg_i dg_i^\prime \varphi(g_1,g_2,g_3) \bar{\varphi}(g_1,g^\prime_2,g^\prime_3) \varphi(g^\prime_1,g^\prime_2,g^\prime_3)\bar{\varphi}(g_1^\prime,g_2,g_3)\,,\label{example}
\end{equation}
assuming that the red edge corresponds to color $1$. As illustrated by the last example in Figure \ref{figBubbles}, graphs can be connected or not, and in this case, they are the product of connected graphs. We call \textit{a bubble} such a connected graph, made of a single piece. Physically, nodes should be interpreted as elementary “atoms of space", and Feynman graphs of the theory reconstruct the cellular decomposition of a discrete space in dimension $d$. 
Because the group is $\U(1)$, the field can be decomposed along group representations $\chi_{\bm p}(\bm \theta):=e^{i \bm p \cdot \bm \theta}$, where $\bm p \in \mathbb{Z}^d$, $ \bm \theta \in ([0,2\pi[\,)^{\times d}$ and $\bm p \cdot \bm \theta$ denotes the standard Euclidean product $\bm p \cdot \bm \theta:= \sum_{c=1}^d p_c \theta_c$. Explicitly:
\begin{equation}
\varphi(\bm g (\bm \theta))=\sum_{\bm p \in \mathbb{Z}^d} \varphi_{\bm p}\, \chi_{\bm p}(\bm \theta)\,.
\end{equation}
It is suitable to work in the Fourier space where the Laplacian is diagonal, and this is indeed what we do for the rest of this paper. A commonly adopted choice for the kinetic part of the classical action is as follows:
\begin{equation}
S_{\text{kin}}[\varphi,\bar{\varphi}]:=\sum_{\bm p} \bar{\varphi}_{\bm p} (\bm p^2+m^2) \Theta^{-1}_a(\Lambda^2-\bm p^2) \varphi_{\bm p}\,,\label{kinaction}
\end{equation}
where the function $\Theta_a$ regularizes UV divergences that may occur in the computation of Feynman amplitudes. Explicitly, we can choose:
\begin{equation}
\Theta_a(x):=\frac{1}{a \sqrt{\pi}} \int_{-\infty}^x\, dy \, e^{-y^2/a^2}\,,
\end{equation}
such that $\lim_{a\to 0}\Theta_a(x)=\Theta(x)$, the ordinary Heaviside step function. Hence, the bare propagator of the theory is diagonal, with entries:
\begin{equation}
C(\bm p):=\frac{\Theta_a(\Lambda^2-\bm p^2)}{\bm p^2+m^2}\,.\label{regularizedC}
\end{equation}
For interactions, the monomials keep the structure detailed in equation \eqref{example}, but integrals are replaced by discrete sums. Hence, a bubble is nothing but a product of the Kronecker delta identifying group field indices accordingly to their respective colors. This choice of regularization \eqref{kinaction} for ultraviolet divergences has been widely considered in the literature \cite{Carrozza_2017a,Lahoche:2018oeo} and reference therein, and is formally equivalent to cutting the degrees of freedom on a ball of radius $\Lambda$, by linking the components together. In the \ref{sectionlack} section of this article, however, we'll use another regularization, on a cube rather than a ball, imposing $p_i\in [-\Lambda,\Lambda]\,\forall i$.\\

Despite the unitary invariance of the interactions, the kinetic kernel $\mathcal{K}$ breaks explicitly the invariance. However the path integral \eqref{partitionfunction} defining $Z[\J,\bar{\J}]$ has to be formally invariant under the transformation:
\begin{equation}
\varphi \to \mathcal{U}_{\bm c}\triangleright \varphi \,,\qquad \bar{\varphi} \to \bar{\varphi}\triangleleft\mathcal{U}_{\bm c}^\dagger\,,
\end{equation}
namely:
\begin{equation}
\delta Z[\J,\bar{\J}]:= \mathcal{U}_{\bm c}\triangleright Z[\J,\bar{\J}] - Z[\J,\bar{\J}] \equiv 0\,. 
\end{equation}
where:
\begin{equation}
\mathcal{U}_{\bm c}\triangleright Z[\J,\bar{\J}]:=\int [d\mathcal{U}_{\bm c}\triangleright\varphi] [d\bar{\varphi}\triangleleft\mathcal{U}_{\bm c}^\dagger] \, \rho[\mathcal{U}_{\bm c}\triangleright\varphi,\bar{\varphi}\triangleleft\mathcal{U}_{\bm c}^\dagger] \exp \left( \bar{\J} \cdot \mathcal{U}_{\bm c}\triangleright\varphi+\bar{\varphi}\triangleleft\mathcal{U}_{\bm c}^\dagger\cdot \J \right)\,.\label{partitionfunction}
\end{equation}
The integration measure is formally invariant by construction, 
\begin{equation}
[d\mathcal{U}_{\bm c}\triangleright\varphi] [d\bar{\varphi}\triangleleft\mathcal{U}_{\bm c}^\dagger]=\det \mathcal{U}_{\bm c} \det \mathcal{U}_{\bm c}^\dagger \,[d\varphi] [d\bar{\varphi}]\equiv [d\varphi] [d\bar{\varphi}]\,.
\end{equation}
Furthermore, the interactions being invariants by construction, 
\begin{equation}
\rho[\mathcal{U}_{\bm c}\triangleright\varphi,\bar{\varphi}\triangleleft\mathcal{U}_{\bm c}^\dagger]= e^{-S_{\text{kin}}[\mathcal{U}_{\bm c}\triangleright\varphi,\bar{\varphi}\triangleleft\mathcal{U}_{\bm c}^\dagger]} e^{-S_{\text{int}}[\varphi,\bar{\varphi}]}\,,
\end{equation}
Finally:
\begin{equation}
0= \int [d\varphi] [d\bar{\varphi}]\,e^{-S_{\text{int}}[\varphi,\bar{\varphi}]} \left(e^{-S_{\text{kin}}[\mathcal{U}_{\bm c}\triangleright\varphi,\bar{\varphi}\triangleleft\mathcal{U}_{\bm c}^\dagger]+\bar{\J} \cdot \mathcal{U}_{\bm c}\triangleright\varphi+\bar{\varphi}\triangleleft\mathcal{U}_{\bm c}^\dagger\cdot \J} - e^{-S_{\text{kin}}[\varphi,\bar{\varphi}]+\bar{\J} \cdot \varphi+\bar{\varphi}\cdot \J}\right) \,.\label{truevariation}
\end{equation}

\subsection{First order Ward identities}

In this section, we briefly recall the first-order Ward identities as studied in the literature. To this end, we consider unitary transformations of $U\in \mathbb{U}(\mathbf{G})$ in the vicinity of the identity, at the linear order with respect to some parameter $\epsilon$:
\begin{equation}
U_{p_1p_2}=\delta_{p_1p_2}+i \epsilon B_{p_1p_2}+\mathcal{O}(\epsilon^2)\,,
\end{equation}
where we work on the Fourier space: $p_1, p_2 \in \mathbb{Z}$, and $B \in \mathfrak{u}(\mathbf{G})$, the Lie algebra of $\mathbb{U}(\mathbf{G})$. Note that because unitary of $U$, we must have $B=B^\dagger$. At order $1$ in $\epsilon$, transformations like \eqref{unitarygen} look like sums over Lie algebra matrices, one for each colors along $\bm c$:
\begin{equation}
\mathcal{U}_{\textbf{c}}(\mathbf{g},\mathbf{g}^\prime):= \prod_{\ell \in \llbracket 1, d \rrbracket^d} \delta(g_\ell^\prime g^{-1}_\ell)+i\epsilon\sum_{c\in {\textbf{c}}}B_c(g_c,g_c^\prime) \prod_{\ell \in \llbracket 1, d \rrbracket^d/c} \delta(g_\ell^\prime g^{-1}_\ell)\,,\label{unitarygen}
\end{equation}
and it is suitable to consider transformations $\mathcal{U}_{\{1\}}$, acting only on the color $c=1$. The variation of the partition function in first order reads:
\begin{equation}
0= \int [d\varphi] [d\bar{\varphi}]\,e^{-S_{\text{int}}[\varphi,\bar{\varphi}]-S_{\text{kin}}[\varphi,\bar{\varphi}]+\bar{\J} \cdot \varphi+\bar{\varphi}\cdot \J} \left(-\delta^{(1)}S_{\text{kin}}+ \delta^{(1)} (\bar{\J} \cdot \varphi+\bar{\varphi}\cdot \J)  \right) + \mathcal{O}(\epsilon^2)\,,\label{variation1storder}
\end{equation}
where $\delta^{(1)}$ denotes variations at first order in $\epsilon$. The variations can be easily computed, and after some algebra, we get:
\bea\label{fin1}
\delta^{(1)} S_{kin}=-i\epsilon\sum_{\bm{p},p_1,p_1^\prime}\delta_{\bm{p}_{\bot_1}\bm{p}\,_{\bot_1}^\prime}\Big(\varphi_{\bm{p}}\big[C^{-1}(\bm{p}\,)-C^{-1}(\bm{p}^\prime)\big]\bar{\varphi}_{\bm{p}^\prime} \Big)B_{p_1^\prime p_1},
\eea
and:
\bea\label{fil}
\delta^{(1)} (\bar{\J} \cdot \varphi+\bar{\varphi}\cdot \J) =i\epsilon \sum_{\bm{p}, p_1^\prime}\Big(\bar{\J }_{p_1^\prime\bm{p}_{\bot_1}}\varphi_{\bm{p}}-\J_{\bm{p}}\,\bar{\varphi}_{p_1^\prime\bm{p}_{\bot_1}} \Big)B_{p_1^\prime p_1}\,,
\eea
where we introduced the notation:
\begin{equation}
\bm{p}_{\bot_i}:=(p_1,\cdots p_{i-1},p_{i+1}\cdots, p_d) \in \mathbb{Z}^{d-1}\,.
\end{equation}
Using these relations, equation \eqref{variation1storder} can be rewritten as:
\begin{equation}
\sum_{p_1,p_1^\prime}B_{p_1^\prime p_1}\sum_{\bm{p}_{\bot_1}}\left( \,\big[C^{-1}(\bm{p}\,)-C^{-1}(\bm{p}^\prime)\big] \frac{\partial^2}{\partial \J_{\bm p^\prime}\partial \bar{\J}_{\bm p}} +  \bar{\J }_{p_1^\prime\bm{p}_{\bot_1}} \frac{\partial}{\partial \bar{\J}_{\bm p}}-\J_{\bm{p}}\,\frac{\partial}{\partial \J_{p_1^\prime\bm{p}_{\bot_1}}}  \right) e^{W[\J,\bar{\J}]}=0\,,
\end{equation}
where we introduced the free energy:
\begin{equation}
W[\J,\bar{\J}]:= \ln Z[\J,\bar{\J}]\,,
\end{equation}
whose derivatives for the sources are $n$-points correlations functions:
\begin{equation}
G^{(n,m)}(\bm p_1,\cdots, \bm p_n; \bm q_1,\cdots, \bm q_m) \equiv \frac{\partial^{n+m} W[\J,\bar{\J}]}{\partial \J_{\bm p_1}\cdots \partial \J_{\bm p_n} \partial \bar{\J}_{\bm q_1}\cdots \partial \bar{\J}_{\bm q_m}}\,.
\end{equation}
Finally, because of the linear independence of the entries of the Lie algebra matrix $B$, 
\begin{equation}\label{Wardone}
\boxed{\sum_{\bm{p}_{\bot_1}}\left( \,\big[C^{-1}(\bm{p}\,)-C^{-1}(\bm{p}^\prime)\big] \frac{\partial^2}{\partial \J_{\bm p^\prime}\partial \bar{\J}_{\bm p}} +  \bar{\J }_{p_1^\prime\bm{p}_{\bot_1}} \frac{\partial}{\partial \bar{\J}_{\bm p}}-\J_{\bm{p}}\,\frac{\partial}{\partial \J_{p_1^\prime\bm{p}_{\bot_1}}}  \right) Z[\J,\bar{\J}]=0\,.}
\end{equation}
This identity provides non-trivial relations between correlations functions of the theory, 
\begin{align}
\nonumber \sum_{\bm{p}_{\bot_1}}\Big( \,\big[C^{-1}(\bm{p}\,)-C^{-1}(\bm{p}^\prime)\big]& \left( G^{(1,1)}(\bm p^\prime,\bm p)+ G^{(1,0)}(\bm p^\prime)G^{(0,1)}(\bm p)\right)
 \\
 &+  \bar{\J }_{p_1^\prime\bm{p}_{\bot_1}} G^{(0,1)}(\bm p)-\J_{\bm{p}}\,G^{(1,0)}(p_1^\prime\bm{p}_{\bot_1}) \Big)=0\,, \label{WIfirst}
\end{align}
where ${\bf p}'=p_1'{\bf p}_{\bot_1}$.
In particular, it provides an interesting relation between the field strength renormalization $Z$ and the quartic coupling constant, of strong interest for renormalization of the just-renormalizable model in rank $d=5$ \cite{Lahoche:2018oeo}. Before considering second-order variations in $\epsilon$, we briefly recall what we expect in ordinary quantum field theory. 

\subsection{Ward identities in quantum electrodynamics}\label{sectionQED}

We close this section with a short review of Ward operator algebra in standard quantum field theory. To remain pedagogical, we consider a concrete example, Euclidean electrodynamics over $\mathbb{R}^D$ with gauge group $\U(1)$, describing interactions between Dirac fermions $\psi(x),\bar{\psi}(x)$. The full partition function of the theory reads:
\begin{equation}
\mathcal{Z}[j,\bar{j},\eta]=\int d[\bar{\psi}] d[\psi]d[A]\,e^{-\int d^Dx \mathcal{L}_m- \int d^Dx \mathcal{L}_A+\int d^Dx (\bar{j}(x) \psi(x)+\bar{\psi}(x) j(x)+\eta(x) A(x))}\,,
\end{equation}
where $\mathcal{L}_m$ and $\mathcal{L}_A$ are respectively lagrangian densities for fermionic matter and gauge field\footnote{Including maybe gauge fixing condition to define properly the path integral.}. Explicitly:
\begin{equation}
\mathcal{L}_m:= \bar{\psi}(x)(i \slashed D-m) \psi(x)\,,
\end{equation}
where: 
\begin{equation}
\slashed D:=\gamma^{\mu}(\partial_\mu -g A_\mu(x))\,,
\end{equation}
$\gamma^{\mu}$ for $\mu=1,\cdots D$ being the standard Dirac matrices and $A_\mu(x)$ is the electromagnetic field. We furthermore denoted as $g$ the coupling constant between matter and gauge field. Note that spinors themselves have group indices. The density lagrangian of the gauge theory is invariant under local gauge transformations:
\begin{equation}
\psi(x) \to e^{i \theta(x)} \psi(x)\,,\qquad \theta(x) \in \mathbb{R}\,,
\end{equation}
provided that $A_\mu(x) \to A_\mu(x)-\partial_\mu \theta(x)$. At the quantum level, the path integral defining the partition function has to be invariant under local transformations $\psi \to e^{i \theta(x)} \psi$, that corresponds to the global translation of the field that remains the formal path integral integration measure invariant. Let us consider an infinitesimal transformation $\psi \to(1+i \theta(x))\psi$. Lagrangian density and source terms change as:
\begin{equation}
\delta \mathcal{L}_m=-\bar{\psi}(x)(\slashed \partial \theta(x))\psi(x)\,,\label{varLm}
\end{equation}
and:
\begin{equation}
\delta S_{\text{sources}}= i\,\int d^Dx (\bar{j}(x) \psi(x)-\bar{\psi}(x) j(x))\theta(x)\,.\label{varJ}
\end{equation}
Hence, because of the formal invariance of the partition function, we get at first order in $\theta(x)$ the Slavnov-Taylor identity:
\begin{equation}
 \left(\partial_\mu  \,  \frac{\partial}{\partial j(x)}\gamma^\mu\frac{\partial}{\partial \bar{j}(x)}+i\,\Big( \bar{j}(x) \frac{\partial}{\partial \bar{j}(x)} -  j(x) \frac{\partial}{\partial j(x)} \Big)\right) e^{W[j,\bar{j},\eta]}\equiv 0\,.
\end{equation}
Let us investigate the contribution of order $\theta^2(x)$ in the computation of the variation of the partition function. The matter lagrangian density remains given by \eqref{varLm}, but the sources variation becomes:
\begin{align}
\delta S_{\text{sources}}&= \,\int d^Dx (\bar{j}(x) \psi(x)-\bar{\psi}(x) j(x))\left(i\theta(x)-\frac{1}{2}\theta^2(x)+\mathcal{O}(\theta^3)\right)\,.\label{varJ2}
\end{align}
The second-order variation reads:
\begin{equation}
\delta^{(2)} \mathcal{Z}=\frac{1}{2}\int d[\bar{\psi}] d[\psi]d[A]\, e^{-S_M} \left( \delta^{(2)} S_{\text{sources}}+\left( - \delta^{(1)} S+ \delta^{(1)} S_{\text{sources}}  \right)^2 \right) \,e^{S_{\text{sources}}}\,,\label{var2}
\end{equation}
where $S_M:=\int d^Dx \mathcal{L}_M$, and $\delta^{(i)}$ designates the variation at order $i$. Let us consider the second contribution, namely:
\begin{align}
\nonumber \delta^{(2)}_2 \mathcal{Z}&:=\frac{1}{2}\int d[\bar{\psi}] d[\psi]d[A]\, e^{-S_M} \left( - \delta^{(1)} S+ \delta^{(1)} S_{\text{sources}}  \right)^2 \,e^{S_{\text{sources}}}\\\nonumber
&= \hat{W}(j,\bar{j}) \, \frac{1}{2}\int d[\bar{\psi}] d[\psi]d[A]\, e^{-S_M} (- \delta^{(1)} S)\,e^{S_{\text{sources}}}\\\nonumber
&+\frac{1}{2}\int d[\bar{\psi}] d[\psi]d[A]\, e^{-S_M} (\delta^{(1)} S_{\text{sources}})\hat{W}(j,\bar{j}) \,e^{S_{\text{sources}}}\,,
\end{align}
where we introduced the Ward operator:
\begin{equation}
\hat{W}(j,\bar{j})\equiv \hat{W}_\partial(j,\bar{j})+\hat{W}_0(j,\bar{j})\,,
\end{equation}
such that:
\begin{equation}
\hat{W}_\partial(j,\bar{j}):=-\int d^Dx\frac{\partial}{\partial j(x)} (\slashed \partial \theta)\frac{\partial}{\partial \bar{j}(x)}\,,
\end{equation}
and:
\begin{equation}
\hat{W}_0(j,\bar{j}):=i\int d^Dx\theta(x)\Big( \bar{j}(x) \frac{\partial}{\partial \bar{j}(x)} -  j(x) \frac{\partial}{\partial j(x)} \Big)\,.
\end{equation}
The second-order variation then becomes:
\begin{equation}
\delta^{(2)}_2 \mathcal{Z}=\frac{1}{2}\hat{W}^2(j,\bar{j})\mathcal{Z}-\frac{1}{2}\int d[\bar{\psi}] d[\psi]d[A]\, e^{-S_M} [\hat{W},\delta^{(1)} S_{\text{sources}})]e^{S_{\text{sources}}}\,. 
\end{equation}
The first term vanishes because of the first order Ward identities. Furthermore, the commutator on the left-hand side can be easily computed and we have:
\begin{equation}
[\hat{W}_\partial,\delta^{(1)} S_{\text{sources}})]e^{S_{\text{sources}}}=0\,,
\end{equation}
\begin{equation}
[\hat{W}_0,\delta^{(1)} S_{\text{sources}})]e^{S_{\text{sources}}}=- \int d^Dx\,\theta^2(x)\Big( \bar{j}(x) \frac{\partial}{\partial \bar{j}(x)} -  j(x) \frac{\partial}{\partial j(x)} \Big) e^{S_{\text{sources}}}\,.
\end{equation}
    Hence, the last term exactly compensates the contribution of $\delta^{(2)} S_{\text{sources}}$ in the expression \eqref{var2}, and $\delta^{(2)} \mathcal{Z}$ vanishes identically taking into account first order Ward identities. There is no additional information taking into account second-order contributions, which is expected because of the group structure of the gauge symmetry: Any finite transformation can be obtained from the composition of a large number of infinitesimal transformations. This illustrates in a simple example what we said in our introduction concerning the redundancy of the Ward identities. As we will see in the next section, this does not work for TGFTs without closure constraints.

\section{Second order Ward identities }\label{secondsection}

In this section, we consider second orders contributions in the $\epsilon$ expansion of identity \eqref{truevariation}, and translate it as a relation between correlation functions as for the first order. Finally, we show that these relations cannot be deduced from first-order Ward identities, and we show that this implies that Ward operators do not compose accordingly with representations of Lie algebra of the group structure, as expected. Furthermore, the derivation as well as the computation given in section \ref{sectionQED} point out the origin of the phenomenon: it is a direct consequence of the non-locality of the interaction. The computation is pedestrian, and we provide the main step for the reader. 

\subsection{Computation at order $\epsilon^2$}
In this section, we compute the second-order Ward identities, i.e. identities coming from order $\epsilon^2$ in the computation of the variation:
\begin{equation}
\mathcal{U}_{\{ab\}}\triangleright Z[\J,\bar{\J}]-Z[\J,\bar{\J}]\equiv 0\,, \label{variationZ}
\end{equation}
where the pair of indices $a,b \in \llbracket 1,d\rrbracket$. The second-order expansions of the unitary matrices $U_c$ are:
\begin{equation}
(U_{c})_{p_1q_1}= \delta_{p_1q_1} + i\epsilon (B_{c})_{p_1q_1}-\frac{1}{2}\epsilon^2\sum_s (B_{c})_{p_1s}(B_{c})_{sq_1}
\end{equation}
\begin{equation}
(\bar{U}_{c})_{p_1q_1}= \delta_{p_1q_1} - i\epsilon(B_{c})_{q_1p_1} - \frac{1}{2}\epsilon^2 \sum_s (B_{c})_{q_1s}(B_{c})_{sp_1}\,.
\end{equation}
We expect that, if $a=b$ in variation \eqref{variationZ}, no violation of the composition hold, because of the internal composition law: Any finite transformation may be obtained as the product of infinitesimal steps. More interesting is the case $a\neq b$, and for simplicity we set $a=1$ and $b=2$, the results obtained in this section remaining valid for arbitrary pairs $(a,b)$. Explicitly, the transformation $\mathcal{U}_{\{12\}}$ reads:
\begin{align}
\nonumber\mathcal{U}_{\{12\}}&=U_{1}\otimes U_{2}\otimes \mathbb{I}^{\otimes (d-2)}=\mathbb{I}^{\otimes d}+ i \bigg(B_1\otimes \mathbb{I}^{\otimes (d-1)} +\mathbb{I}\otimes B_2\otimes \mathbb{I}^{\otimes (d-2)}\bigg)\\
&-\frac{1}{2}\bigg(B_1^2\otimes \mathbb{I}^{\otimes (d-1)}  + \mathbb{I} \otimes B_{2}^2 \otimes \mathbb{I}^{\otimes (d-2)}  \bigg) - B_{1}\otimes B_{2}\otimes \mathbb{I}^{\otimes (d-2)}\,,\label{gladiateur}
\end{align}
where, to simplify the notations, we assume $B_a=\mathcal{O}(\epsilon)$ $\forall a$. Furthermore, because of the formal invariance of the path integration measure and the interactions, we have:
\begin{align}
\nonumber \mathcal{U}_{\{12\}}\triangleright Z[\J,\bar{\J}] = \int [d\varphi] [d\bar{\varphi}] &\exp \Bigg(- S_{\text{kin}}[\mathcal{U}_{\{12\}}\triangleright \varphi,\bar{\varphi}\triangleleft \mathcal{U}_{\{12\}}^\dagger]-S_{\text{int}}[\varphi,\bar{\varphi}] \\
&+\left( \bar{\J} \cdot \mathcal{U}_{\{12\}}\triangleright\varphi+\bar{\varphi}\triangleleft\mathcal{U}_{\{12\}}^\dagger\cdot \J \right)\Bigg)\,.
\end{align}
The first piece that we have to compute is the variation of the kinetic action, which reads explicitly as:
\begin{align}
\nonumber \Delta_{12}S_{\text{kin}}[\varphi,\bar{\varphi}]:&=S_{\text{kin}}[\mathcal{U}_{\{12\}}\triangleright \varphi,\bar{\varphi}\triangleleft \mathcal{U}_{\{12\}}^\dagger] - S_{\text{kin}}[\varphi,\bar{\varphi}]\\
&=\sum_{\bm p}\bigg[\bar{\varphi}\triangleleft \mathcal{U}_{\{12\}}^\dagger\bigg]_{\bm p} C^{-1}(\bm p)\bigg[\mathcal{U}_{\{12\}}\triangleright \varphi\bigg]_{\bm p}-\sum_{\bm p}\bar{\varphi}_{\bm p} C^{-1}(\bm p)\varphi_{\bm p}\,,
\end{align}
where the bare propagator should include regularization, see \eqref{regularizedC}. The action of the unitary transform is explicitly given by:
\begin{equation}
\bigg[\mathcal{U}_{\{12\}}\triangleright \varphi\bigg]_{\bm{p}} = \sum_{q_1q_2} (U_{1})_{p_1q_1}(U_{2})_{p_2q_2} \varphi_{q_1q_2\bm{p}_{\bot \bot}}\,,
\end{equation}
where we introduced the notation:
\begin{equation}
\bm{p}_{\bot \bot}:=(p_3,\cdots,p_d) \in \mathbb{Z}^{d-2}\,.
\end{equation}
Therefore:
\begin{align}
\nonumber &\Delta_{12}S_{\text{kin}}[\varphi,\bar{\varphi}]= \sum_{\bm{p}\in \mathbb{Z}^d} \sum_{q_1q_2} \bigg[\delta^{(1)}_{p_1q_1}\delta^{(2)}_{p_2q_2}+i\bigg( \delta_{p_2q_2}(B_{1})_{p_1q_1} + \delta_{p_1q_1}(B_{2})_{p_2q_2}\bigg)\\\nonumber
&- \frac{1}{2} \bigg(\delta_{p_2q_2}  (B_{1}^2)_{p_1q_1}  + 
 \delta_{p_1q_1} (B_{2}^2)_{p_2 q_2}\bigg) - B^{(1)}_{p_1q_1}B^{(2)}_{p_2q_2} \bigg] \varphi_{q_1q_2\bm{p}_{\bot \bot}} C^{-1}(\bm{p}) \sum_{r_1r_2}\bigg[ \delta_{p_1r_1}\delta_{p_2r_2}\\\nonumber
 &- i\bigg( \delta_{p_2r_2}(B_1)_{r_1p_1} + \delta_{p_1r_1}(B_2)_{r_2p_2}\bigg) - \frac{1}{2} \bigg(\delta_{p_2r_2} (B_{1}^2)_{r_1p_1} \\
 &+ \delta_{p_1r_1} (B_2^2)_{r_2p_2}\bigg)- B^{(1)}_{r_1p_1}B^{(2)}_{r_2p_2} \bigg]\bar{\varphi}_{r_1r_2\bm{p}_{\bot \bot}}-S_{\text{kin}}[\varphi,\bar{\varphi}]+\mathcal{O}(\epsilon^3)\,.
\end{align}
This expression can be expanded in the power of $\epsilon$, and identifying contributions for $B_1$ and $B_2$, we have for the first order nothing but what we obtained in the previous section:
\begin{equation}
\Delta_{12}S_{\text{kin}}\bigg|_{\mathcal{O}(B_{1})}
= -i \sum_{\bm{p}_{\bot_1}} \sum_{q_1r_1} \varphi_{q_1\bm{p}_{\bot_1}} \bigg( C^{-1}(q_1\bm{p}_{\bot_1}) - C^{-1}(r_1\bm{p}_{\bot_1}) \bigg)\bar{\varphi}_{r_1\bm{p}_{\bot_1}}  (B_{1})_{r_1q_1}\,,
\end{equation}
and
\begin{equation}
\Delta_{12}S_{\text{kin}}\bigg|_{\mathcal{O}(B_{2})}
= -i \sum_{\bm{p}_{\bot_2}} \sum_{q_2r_2} \varphi_{q_2\bm{p}_{\bot_2}} \bigg( C^{-1}(q_2\bm{p}_{\bot_2}) - C^{-1}(r_2\bm{p}_{\bot_2}) \bigg)\bar{\varphi}_{r_2\bm{p}_{\bot_2}}  (B_{2})_{r_2q_2}\,,
\end{equation}
where we used of the obvious notation $q_1\bm{p}_{\bot_1}\equiv (q_1,\bm{p}_{\bot_1})\in \mathbb{Z}$. 
The second order contributions are of two kinds: we have terms of order $\mathcal{O}(B_a^2)$, for $a=1,2$, and of order $\mathcal{O}(B_1B_2)$. Explicitly:
\begin{equation}
\Delta_{12}S_{\text{kin}}\bigg|_{\mathcal{O}(B_{1}^2)}=\sum_{\bm{p}_{\bot_1}} \sum_{q_{1}r_{1}s}\varphi_{q_1\vec{p}_{\bot_1}}\bigg(C^{-1}(s\vec{p}_{\bot_1}) - \frac{1}{2}C^{-1}(q_1\vec{p}_{\bot_1}) -\frac{1}{2}C^{-1}(r_1\vec{p}_{\bot_1}) \bigg) \bar{\varphi}_{r_1\vec{p}_{\bot_1}} (B_{1})_{r_1s}(B_{1})_{sq_1}\,,\label{OB12}
\end{equation}
and the same thing for $\mathcal{O}(B_2^2)$, and for the mixed contribution:
\begin{align}
\nonumber \Delta_{12}S_{\text{kin}}\bigg|_{\mathcal{O}(B_{1}B_2)}&= \sum_{\bm{p}} \sum_{r_1,r_2} \varphi_{\bm{p}} \bigg( C^{-1}(\bm{p}) +C^{-1}(r_1r_2 \bm{p}_{\bot \bot}) \\
&- C^{-1}(r_1 \bm{p}_{\bot_1}) - C^{-1}(r_2\bm{p}_{\bot_2})\bigg) \bar{\varphi}_{r_1r_2 \bm{p}_{\bot \bot}} (B_{1})_{r_1p_1} (B_{2})_{r_2p_2}\,.
\end{align}
The first and second-order variations for the source term can be computed in the same way. We have:
\begin{equation}
\Delta_{12}S_{\text{source}}:=\left( \bar{\J} \cdot \mathcal{U}_{\{12\}}\triangleright\varphi+\bar{\varphi}\triangleleft\mathcal{U}_{\{12\}}^\dagger\cdot \J \right)-\left( \bar{\J} \cdot \varphi+\bar{\varphi}\cdot \J \right)\,,
\end{equation}
and a tedious calculation for order $1$ ad order $2$ contributions leads to:
\begin{equation}
\Delta_{12}S_{\text{source}}\bigg|_{\mathcal{O}(B_{2})}=i\sum_{\vec{p}_{\bot_2},r_2,q_2}\Bigg(\bar{\J}_{\vec{p}}\,\varphi_{q_2\vec{p}_{\bot_2}}-\J_{\vec{p}}\,\bar{\varphi}_{r_2\vec{p}_{\bot_2}}\Bigg)(B_2)_{r_2q_2}\,,
\end{equation}
and:
\begin{equation}
\Delta_{12}S_{\text{source}}\bigg|_{\mathcal{O}(B_{1}^2)}=-\dfrac{1}{2}\sum_{\bm{p}_{\bot_1},r_1,q_1}\Big(\bar{\J}_{r_1\bm{p}_{\bot_1}}\varphi_{q_1\bm{p}_{\bot_1}}+\J_{q_1\bm{p}_{\bot_1}}\bar{\varphi}_{r_1\bm{p}_{\bot_1}}\Big)(B_{1}^2)_{r_1q_1}\,,
\end{equation}
\begin{align}
\Delta_{12}S_{\text{source}}\bigg|_{\mathcal{O}(B_{1}B_2)}=-\sum_{\bm{p},r_1,r_2}\Big(\bar{\J}_{r_1r_2\bm{p}_{\bot\bot}}\,\varphi_{\vec{p}}-\J_{\bm{p}}\,
\bar{\varphi}_{r_1r_2\bm{p}_{\bot\bot}}\Big)(B_{1})_{r_1p_1}(B_{2})_{r_2p_2}\,.
\end{align}
To summarize, we have:
\bea
\Delta_{12}S_{\text{kin}}=:B_1x_1+B_2x_2+B_1x_{11}B_1+B_{2}x_{22}B_{2}+B_{1}x_{12}B_{2}\,,
\eea
\bea
\Delta_{12}S_{\text{source}}=:B_{1}y_1+B_{2}y_2+B_{1}y_{11}B_{1}+B_{2}y_{22}B_{2}+B_{1}y_{12}B_{2}\,,
\eea
where the matrix-valued coefficients $x_i$, $x_{ij}$, $y_i$ and $y_{ij}$ are defined by the previous equations, each monomial being indeed a contraction overall indices of matrices $B$, for instance:
\begin{equation}
B_1x_1:=\sum_{a,b} (B_1)_{ab}(x_1)_{ab}\,.
\end{equation}
Using these notations, we can compute the variation of the partition function
\begin{align}
\Delta_{12} Z[\J,\bar{\J}] :=\mathcal{U}_{\{12\}}\triangleright Z[\J,\bar{\J}] -Z[\J,\bar{\J}]\,.
\end{align}
Explicitly:
\begin{align}
\nonumber\Delta_{12} Z[\J,\bar{\J}]=:\int [d\varphi]\,[d\bar{\varphi}]\,e^{-S+S_{\text{source}}}\times \Big[X+\dfrac{1}{2}Y\Big].
\end{align}
where:
\begin{equation}
Y:=\Big(B_{1}(-x_1+y_1)+B_{2}(-x_2+y_2)\Big)^2\,,
\end{equation}
and:
\begin{align}
\nonumber X&:=B_{1}(-x_1+y_1)+B_{2}(-x_2+y_2)+B_{1}(-x_{11}+y_{22})B_{1}\\
&+B_{2}(-x_{22}+y_{22})B_{2}+B_{1}(-x_{12}+y_{12})B_{2}\,.
\end{align}
Let us investigate the term of order $(B_1)^2$. The contributions are of two kinds. We have contributions involving the matrix product $(B_1^2)_{ab}:=\sum_s (B_1)_{as}(B_1)_{sb}$ (see equation \eqref{OB12}). This corresponds to terms of the form $B_1x_{11}B_1$ and $B_1y_{11}B_1$ in the variation of the partition function. On the other hand, we have terms involving Kronecker products, coming from contributions like $(B_1x_{1})^2$. To summary, the contribution of order $(B_1)^2$ is of the form:
\begin{equation}
\Delta_{12} Z[\J,\bar{\J}]\bigg\vert_{\mathcal{O}(B_1^2)}=: \sum_{q_1,r_1,s}\, (W_{11})_{q_1r_1s} (B_1)_{r_1 s} (B_1)_{sq_1}+\sum_{q_1,r_1,s\neq s^\prime} (\tilde{W}_{11})_{q_1r_1ss^{\prime}}(B_1)_{r_1 s} (B_1)_{s^\prime q_1}\,,
\end{equation}
where:
\begin{align}
W_{11}&=\sum_{\bm{p}_{\bot_1}}\int [d\varphi]\,[d\bar{\varphi}]\,e^{-S+S_{\text{source}}}\Bigg[\varphi_{q_1\bm{p}_{\bot_1}}\Big(C^{-1}(s\bm{p}_{\bot_1})-\dfrac{1}{2}C^{-1}(q_1\bm{p}_{\bot_1})-\dfrac{1}{2}C^{-1}(r_1\bm{p}_{\bot_1}) \Big)\bar{\varphi}_{r_1\bm{p}_{\bot_1}}\cr
&+\dfrac{1}{2}\sum_{\bm{p}_{\bot_1},r_1,q_1,s}\Big(\bar{\J}_{r_1\bm{p}_{\bot_1}}\varphi_{q_1\bm{p}_{\bot_1}}+\J_{q_1\bm{p}_{\bot_1}}\bar{\phi}_{r_1\bm{p}_{\bot_1}}\Big)\cr
&+\frac{1}{2}\varphi_{s\bm{p}_{\bot_1}}\varphi_{q_1\bm{p}_{\bot_1}}\Big(C^{-1}(s\bm{p}_{\bot_1})-C^{-1}(r_1\bm{p}_{\bot_1}) \Big)\Big(C^{-1}(q_1\bm{p}_{\bot_1})-C^{-1}(s\bm{p}_{\bot_1}) \Big)\bar{\varphi}_{s\bm{p}_{\bot_1}}\bar{\varphi}_{r_1\bm{p}_{\bot_1}}\cr
&+\frac{1}{2}\varphi_{s\bm{p}_{\bot_1}}\Big(C^{-1}(s\bm{p}_{\bot_1})-C^{-1}(r_1\bm{p}_{\bot_1}) \Big)\bar{\varphi}_{r_1\bm{p}_{\bot_1}}\Big(\bar{\J}_{\bm{p}}\,\varphi_{q_1\bm{p}_{\bot_1}} -\J_{\bm{p}}\,\bar{\varphi}_{s\bm{p}_{\bot_1}}\Big)\cr
&+\frac{1}{2}\varphi_{q_1\bm{p}_{\bot_1}}\Big(C^{-1}(q_1\bm{p}_{\bot_1})-C^{-1}(s\bm{p}_{\bot_1}) \Big)\bar{\varphi}_{s\bm{p}_{\bot_1}}\Big(\bar{\J}_{\bm{p}}\,\varphi_{s\bm{p}_{\bot_1}} -\J_{\bm{p}}\,\bar{\varphi}_{r_1\bm{p}_{\bot_1}}\Big)\cr
&+\frac{1}{2}\Big(\bar{\J}_{\bm{p}}\,\varphi_{s\bm{p}_{\bot_1}} -\J_{\bm{p}}\,\bar{\varphi}_{r_1\bm{p}_{\bot_1}}\Big)\Big(\bar{\J}_{\bm{p}}\,\varphi_{q_1\bm{p}_{\bot_1}} -\J_{\bm{p}}\,\bar{\varphi}_{s\bm{p}_{\bot_1}}\Big)\Bigg]=0\,,
\end{align}
\begin{align}
&\widetilde{W}_{11}=\sum_{\bm{p}_{\bot_1}}\int [d\varphi]\,[d\bar{\varphi}]\,e^{-S+S_{\text{source}}}\Bigg[\varphi_{s\bm{p}_{\bot_1}}\varphi_{q_1\bm{p}_{\bot_1}}\Big(C^{-1}(s\bm{p}_{\bot_1})-C^{-1}(r_1\bm{p}_{\bot_1}) \Big)\\\nonumber
&\times\Big(C^{-1}(q_1\bm{p}_{\bot_1})-C^{-1}(s'\bm{p}_{\bot_1}) \Big)\bar{\varphi}_{s'\bm{p}_{\bot_1}}\bar{\varphi}_{r_1\bm{p}_{\bot_1}}+\varphi_{s\bm{p}_{\bot_1}}\Big(C^{-1}(s\bm{p}_{\bot_1})-C^{-1}(r_1\bm{p}_{\bot_1}) \Big)\cr
&\times \bar{\varphi}_{r_1\bm{p}_{\bot_1}}\Big(\bar{\J}_{\bm{p}}\,\varphi_{q_1\bm{p}_{\bot_1}} -\J_{\bm{p}}\,\bar{\varphi}_{s'\bm{p}_{\bot_1}}\Big)+\Big(\bar{\J}_{\bm{p}}\,\phi_{s\bm{p}_{\bot_1}} -\J_{\bm{p}}\,\bar{\varphi}_{r_1\bm{p}_{\bot_1}}\Big)\varphi_{q_1\bm{p}_{\bot_1}}\Big(C^{-1}(q_1\bm{p}_{\bot_1})-C^{-1}(s'\bm{p}_{\bot_1}) \Big)\bar{\varphi}_{s'\bm{p}_{\bot_1}}\cr
&+\Big(\bar{\J}_{\bm{p}}\,\varphi_{q_1\bm{p}_{\bot_1}} -\J_{\bm{p}}\,\bar{\phi}_{s'\bm{p}_{\bot_1}}\Big)\Big(\bar{\J}_{\bm{p}}\,\varphi_{s\bm{p}_{\bot_1}} -\J_{\bm{p}}\,\bar{\varphi}_{r_1\bm{p}_{\bot_1}}\Big)\Bigg] = 0\,.
\end{align}
In the same way, we have coefficients $W_{22}$ and $\widetilde{W}_{22}$ for variation of order $B_2^2$, which can be deduced from the previous expressions. In the same way, the variation of order $B_1B_2$ takes the form:
\begin{equation}
\Delta_{12} Z[\J,\bar{\J}]\bigg\vert_{\mathcal{O}(B_1B_2)}=: \sum_{r_1,p_1,r_2,p_2}\, (W_{12})_{p_1p_2r_1r_2} (B_1)_{r_1 p_1} (B_2)_{r_2p_2}\,,
\end{equation}
where:
\begin{align}
&W_{12} = \sum_{\bm{p}_{\bot\bot}} \int [d\varphi]\,[d\bar{\varphi}]\,e^{-S+S_{\text{source}}}\bigg[ \varphi_{\bm{p}} \bigg(C^{-1}(\bm{p})  + C^{-1}(r_1r_2 \bm{p}_{\bot \bot})- C^{-1}(r_1\bm{p_{\bot_1}})  - C^{-1}(r_2 \bm{p}_{\bot_2})\bigg)\bar{\varphi}_{r_1r_2 \bm{p}_{\bot \bot}} \cr
&- \bigg( \bar{\J}_{r_1r_2 \bm{p}_{\bot \bot}}\varphi_{\bm{p}} - \J_{\bm{p}} \bar{\varphi}_{r_1r_2 \bm{p}_{\bot \bot}}\bigg)-\varphi_{\bm{p}}\varphi_{\bm{p}}\bigg( C^{-1}(\bm{p})  - C^{-1}(r_1 \bm{p}_{\bot_1})\bigg)\bigg( C^{-1}(\bm{p})  - C^{-1}(r_2 \bm{p}_{\bot_2})\bigg)\bar{\varphi}_{r_1 \bm{p}_{\bot_1}}\bar{\varphi}_{r_2 \bm{p}_{\bot_2}} \cr
 &-\varphi_{\bm{p}}\bigg( C^{-1}(\bm{p})  - C^{-1}(r_1 \bm{p}_{\bot_1})\bigg)\bar{\varphi}_{r_1 \bm{p}_{\bot_1}} \bigg( \bar{\J}_{ \bm{p}}\varphi_{\bm{p}} - \J_{\bm{p}} \bar{\varphi}_{r_2 \bm{p}_{\bot_2}} \bigg) \cr
 &-\bigg( \bar{\J}_{ \bm{p}}\varphi_{\bm{p}} - \J_{\bm{p}} \bar{\varphi}_{r_1 \bm{p}_{\bot_1}} \bigg)\varphi_{\bm{p}}\bigg( C^{-1}(\bm{p})  - C^{-1}(r_2 \bm{p}_{\bot_2})\bigg)\bar{\varphi}_{r_2 \bm{p}_{\bot_2}} -\bigg( \bar{\J}_{ \bm{p}}\varphi_{\bm{p}} - \J_{\bm{p}} \bar{\varphi}_{r_1 \bm{p}_{\bot_1}} \bigg)\bigg( \bar{\J}_{ \bm{p}}\varphi_{\bm{p}} - \J_{\bm{p}} \bar{\varphi}_{r_2 \bm{p}_{\bot_2}} \bigg) \bigg] \,.
\end{align}

It is suitable, for the discussion of the next section, to express the coefficients $W_{ij}$ in terms of observables. After a tedious calculation but without technical difficulties, we arrive at the following expressions for $W_{11}$:
\bea\label{machiavel}
W_{11}=:\sum_{{\bf p}_{\bot_1}}\Big( \Delta_{C_1}\cI_1+\Delta_{C_2}\cI_2+\Delta_{C_3}\cI_3+\Delta_{C_4}\cI_4+\mathcal{D}\Big)
 \eea
where:
\bea
\mathcal{I}_1&=&\bigg(\frac{\partial W}{\partial \bar{\J}_{q_1\bm{p}_{\bot_1}}}\frac{\partial W}{\partial \J_{r_1\bm{p}_{\bot_1}}}+\frac{\partial^2 W}{\partial \J_{r_1\bm{p}_{\bot_1}}\partial\bar{\J}_{q_1\bm{p}_{\bot_1}}} \bigg)
\eea
\bea
\mathcal{I}_2&=&\frac{1}{2}\bigg\{ \frac{\partial W}{\partial \J_{s\bm{p}_{\bot_1}}}\frac{\partial^2 W}{\partial \bar{\J}_{q_1\bm{p}_{\bot_1}}\partial \J_{r_1\bm{p}_{\bot_1}}}\frac{\partial W}{\partial \bar{\J}_{s\bm{p}_{\bot_1}}} + \frac{\partial^3 W}{\partial \bar{\J}_{q_1\bm{p}_{\bot_1}}\partial \J_{s\bm{p}_{\bot_1}}\partial \J_{r_1\bm{p}_{\bot_1}}}\frac{\partial W}{\partial \bar{\J}_{s\bm{p}_{\bot_1}}}\cr 
&+&\frac{\partial W}{\partial \J_{s\bm{p}_{\bot_1}}}\frac{\partial W}{\partial \bar{\J}_{q_1\bm{p}_{\bot_1}}}\frac{\partial^2 W}{\partial \bar{\J}_{s\bm{p}_{\bot_1}}\partial \J_{r_1\bm{p}_{\bot_1}}}+ \frac{\partial^2 W}{\partial \bar{\J}_{q_1\bm{p}_{\bot_1}}\partial \J_{s\bm{p}_{\bot_1}}}\frac{\partial^2 W}{\partial \bar{\J}_{s\bm{p}_{\bot_1}}\partial \J_{r_1\bm{p}_{\bot_1}}} \cr
&+& \frac{\partial^2 W}{\partial \bar{\J}_{q_1\bm{p}_{\bot_1}}\partial \J_{r_1\bm{p}_{\bot_1}}}\frac{\partial^2 W}{\partial \bar{\J}_{s\bm{p}_{\bot_1}}\partial \J_{s\bm{p}_{\bot_1}}} 
+ \frac{\partial W}{\partial \bar{\J}_{q_1\bm{p}_{\bot_1}}}\frac{\partial^3 W}{\partial \bar{\J}_{s\bm{p}_{\bot_1}}\partial \J_{s\bm{p}_{\bot_1}}\partial \J_{r_1\bm{p}_{\bot_1}}}\cr
&+&\frac{\partial^2 W}{\partial \J_{s\bm{p}_{\bot_1}}\partial \J_{r_1\bm{p}_{\bot_1}}}\bigg( \frac{\partial W}{\partial \bar{\J}_{q_1\bm{p}_{\bot_1}}}\frac{\partial W}{\partial \bar{\J}_{s\bm{p}_{\bot_1}}} + \frac{\partial^2 W}{\partial \bar{\J}_{s\bm{p}_{\bot_1}}\partial \bar{\J}_{q_1\bm{p}_{\bot_1}}} \bigg) 
+ \frac{\partial W}{\partial  \J_{q_1\bm{p}_{\bot_1}}}\frac{\partial^3 W}{\partial \bar{\J}_{s\bm{p}_{\bot_1}}\partial \bar{\J}_{q_1\bm{p}_{\bot_1}}\J_{r_1\bm{p}_{\bot_1}}}\cr
&+& \frac{\partial W}{\partial  \J_{r_1\bm{p}_{\bot_1}}} \bigg[\frac{\partial^2 W}{\partial \bar{\J}_{q_1\bm{p}_{\bot_1}}\partial \J_{s\bm{p}_{\bot_1}}}\frac{\partial W}{\partial  \bar{\J}_{s\bm{p}_{\bot_1}}}  +\frac{\partial W}{\partial  \bar{\J}_{q_1\bm{p}_{\bot_1}}}\frac{\partial^2 W}{\partial \bar{\J}_{s\bm{p}_{\bot_1}}\partial \J_{s\bm{p}_{\bot_1}}}\cr 
&+& \frac{\partial W}{\partial  \J_{s\bm{p}_{\bot_1}}} \bigg(\frac{\partial W}{\partial  \bar{\J}_{q_1\bm{p}_{\bot_1}}}\frac{\partial W}{\partial  \bar{\J}_{s\bm{p}_{\bot_1}}} 
+ \frac{\partial^2 W}{\partial  \bar{\J}_{s\bm{p}_{\bot_1}}\partial  \bar{\J}_{q_1\bm{p}_{\bot_1}}}  \bigg) +  \frac{\partial^3 W}{\partial  \bar{\J}_{s\bm{p}_{\bot_1}}\partial  \bar{\J}_{q_1}\bm{p}_{\bot_1}\partial  \J_{s\bm{p}_{\bot_1}}} \bigg]\cr 
	&+& \frac{\partial^4 W}{\partial  \bar{\J}_{s\bm{p}_{\bot_1}}\partial  \bar{\J}_{q_1\bm{p}_{\bot_1}}\partial  \J_{s\bm{p}_{\bot_1}}\partial  \J_{r_1\bm{p}_{\bot_1}}}\bigg\}
\eea
\bea
\mathcal{I}_3&=&-\frac{1}{2}\bigg\{ \bar{\J}_{\bm{p}} \bigg[ \frac{\partial^2 W}{\partial  \bar{\J}_{q_1\bm{p}_{\bot_1}}\partial  \J_{r_1\bm{p}_{\bot_1}}}\frac{\partial W}{\partial  \bar{\J}_{s\bm{p}_{\bot_1}}} +\frac{\partial W}{\partial  \bar{\J}_{q_1\bm{p}_{\bot_1}}}\frac{\partial^2 W}{\partial  \bar{\J}_{s\bm{p}_{\bot_1}}\partial  \J_{r_1\bm{p}_{\bot_1}}}\cr 
&+&\frac{\partial W}{\partial  \J_{r_1\bm{p}_{\bot_1}}}\bigg(\frac{\partial W}{\partial  \bar{\J}_{q_1\bm{p}_{\bot_1}}}\frac{\partial W}{\partial  \bar{\J}_{s\bm{p}_{\bot_1}}} +  \frac{\partial^2 W}{\partial  \bar{\J}_{s\bm{p}_{\bot_1}}\partial  \bar{\J}_{q_1\bm{p}_{\bot_1}}}\bigg) + \frac{\partial^3 W}{\partial \bar{\J}_{s\bm{p}_{\bot_1}}\partial \bar{\J}_{q_1\bm{p}_{\bot_1}}\partial \J_{r_1\bm{p}_{\bot_1}}} \bigg]\cr
&-& \J_{\bm{p}} \bigg[  \frac{\partial^2 W}{\partial  \J_{r_1\bm{p}_{\bot_1}}\partial  \J_{s\bm{p}_{\bot_1}}}\frac{\partial W}{\partial  \bar{\J}_{s\bm{p}_{\bot_1}}} 
+ \frac{\partial W}{\partial  \J_{r_1\bm{p}_{\bot_1}}}\frac{\partial^2 W}{\partial  \bar{\J}_{s\bm{p}_{\bot_1}}\partial  \J_{s\bm{p}_{\bot_1}}} \cr
&+& \frac{\partial W}{\partial  \J_{s\bm{p}_{\bot_1}}}\bigg(\frac{\partial W}{\partial  \J_{r_1\bm{p}_{\bot_1}}}\frac{\partial W}{\partial  \bar{\J}_{s\bm{p}_{\bot_1}}} + \frac{\partial^2}{\partial  \bar{\J}_{s\bm{p}_{\bot_1}}\partial  \J_{r_1\bm{p}_{\bot_1}}}\bigg) + \frac{\partial^3 W}{\partial \bar{\J}_{s\bm{p}_{\bot_1}}\partial \J_{r_1\bm{p}_{\bot_1}}\partial \J_{s\bm{p}_{\bot_1}}} \bigg]\bigg\}
\eea
\bea
\mathcal{I}_4&=&-\frac{1}{2}\bigg\{\bar{\J}_{\bm{p}} \bigg[  \frac{\partial^2 W}{\partial  \bar{\J}_{s\bm{p}_{\bot_1}}\partial  \J_{s\bm{p}_{\bot_1}}}\frac{\partial W}{\partial  \bar{\J}_{q_1\bm{p}_{\bot_1}}} +\frac{\partial}{\partial  \bar{\J}_{s\bm{p}_{\bot_1}}}\frac{\partial^2 W}{\partial  \bar{\J}_{q_1\bm{p}_{\bot_1}}\partial  \J_{s\bm{p}_{\bot_1}}} \cr
&+&\frac{\partial W}{\partial  \J_{s\bm{p}_{\bot_1}}}\bigg(\frac{\partial W}{\partial  \bar{\J}_{s\bm{p}_{\bot_1}}}\frac{\partial W}{\partial  \bar{\J}_{q_1\bm{p}_{\bot_1}}} 
	+  \frac{\partial^2 W}{\partial  \bar{\J}_{q_1\bm{p}_{\bot_1}}\partial  \bar{\J}_{s\bm{p}_{\bot_1}}}\bigg) + \frac{\partial^3 W}{\partial \bar{\J}_{q_1\bm{p}_{\bot_1}}\partial \bar{\J}_{s\bm{p}_{\bot_1}}\partial \J_{s\bm{p}_{\bot_1}}} \bigg]  \cr
 &-& \J_{\bm{p}} \bigg[  \frac{\partial^2 W}{\partial  \J_{s\bm{p}_{\bot_1}}\partial  \J_{r_1\bm{p}_{\bot_1}}}\frac{\partial W}{\partial  \bar{\J}_{q_1\bm{p}_{\bot_1}}} +\frac{\partial W}{\partial  \J_{s\bm{p}_{\bot_1}}}\frac{\partial^2 W}{\partial  \bar{\J}_{q_1\bm{p}_{\bot_1}}\partial  \J_{r_1\bm{p}_{\bot_1}}}  \cr 
	&+& \frac{\partial W}{\partial  \J_{r_1\bm{p}_{\bot_1}}}\bigg(\frac{\partial W}{\partial  \J_{s\bm{p}_{\bot_1}}}\frac{\partial W}{\partial  \bar{J}_{q_1\bm{p}_{\bot_1}}} + \frac{\partial^2 W}{\partial  \bar{J}_{q_1\bm{p}_{\bot_1}}\partial  \J_{s\bm{p}_{\bot_1}}}\bigg) + \frac{\partial^3 W}{\partial \bar{\J}_{q_1\bm{p}_{\bot_1}}\partial \J_{s\bm{p}_{\bot_1}}\partial \J_{r_1\bm{p}_{\bot_1}}} \bigg]\bigg\} 
\eea
\bea
\mathcal{D}&=&\frac{1}{2} \bigg(\bar{\J}_{r_1\bm{p}_{\bot_1}}\frac{\partial W}{\partial \bar{\J}_{q_1\bm{p}_{\bot_1}}} +\J_{q_1\bm{p}_{\bot_1}} \frac{\partial W}{\partial \J_{r_1\bm{p}_{\bot_1}}} \bigg)
 \cr
 &+&  \frac{1}{2} \bigg[\bar{\J}_{\bm{p}}\bar{\J}_{\bm{p}}\bigg(\frac{\partial W}{\partial  \bar{\J}_{s\bm{p}\bot_1}}\frac{\partial W}{\partial  \bar{\J}_{q_1\bm{p}_{\bot_1}}}+  \frac{\partial^2W}{\partial  \bar{\J}_{s\bm{p}_{\bot_1}}\partial  \bar{\J}_{q_1\bm{p}_{\bot_1}}} \bigg) \cr
 &-& \bar{\J}_{\bm{p}}\J_{\bm{p}}\bigg(\frac{\partial W}{\partial  \bar{\J}_{s\bm{p}\bot_1}}\frac{\partial W}{\partial  \J_{s\bm{p}_{\bot_1}}}+  \frac{\partial^2 W}{\partial  \bar{\J}_{s\bm{p}_{\bot_1}}\partial  \J_{s\bm{p}_{\bot_1}}} \bigg) \cr 
	&-& \J_{\bm{p}}\bar{\J}_{\bm{p}}\bigg(\frac{\partial W}{\partial  \J_{r_1\bm{p}\bot_1}}\frac{\partial W}{\partial  \bar{\J}_{q_1\bm{p}_{\bot_1}}}+  \frac{\partial^2 W}{\partial  \J_{r_1\bm{p}_{\bot_1}}\partial  \bar{J}_{q_1\bm{p}_{\bot_1}}} \bigg) \cr
 &+& \J_{\bm{p}}\J_{\bm{p}}\bigg(\frac{\partial W}{\partial  \J_{r_1\bm{p}\bot_1}}\frac{\partial W}{\partial  \J_{s\bm{p}_{\bot_1}}}+  \frac{\partial^2 W}{\partial  \J_{r_1\bm{p}_{\bot_1}}\partial  \J_{s\bm{p}_{\bot_1}}} \bigg)\bigg]
\eea
with:
	\bea
	\Delta_{C_1}&=&\Big(C^{-1}(s\bm{p}_{\bot_1})-\dfrac{1}{2}C^{-1}(q_1\bm{p}_{\bot_1})-\dfrac{1}{2}C^{-1}(r_1\bm{p}_{\bot_1}) \Big)\cr
	\Delta_{C_2}&=&\Big(C^{-1}(s\bm{p}_{\bot_1})-C^{-1}(r_1\bm{p}_{\bot_1}) \Big)\Big(C^{-1}(q_1\bm{p}_{\bot_1})-C^{-1}(s\bm{p}_{\bot_1}) \Big)\cr
	\Delta_{C_3}&=&\Big(C^{-1}(r_1\bm{p}_{\bot_1})-C^{-1}(s\bm{p}_{\bot_1}) \Big)\cr
	\Delta_{C_4}&=&\Big(C^{-1}(s\bm{p}_{\bot_1})-C^{-1}(q_1\bm{p}_{\bot_1}) \Big)\,,
	\eea
The other components $W_{ij}$ and $\tilde{W}_{ii}$ could be computed explicitly as well, their explicit form being irrelevant to the argument presented in the next section.

\subsection{Lack of redundancy for second order Ward identities}\label{sectionlack}

In this section, we show finally that second-order Ward identities cannot be reduced to first-order ones, and thus provides us additional relations between observable. Let us consider the first-order WT identity given by the relation \eqref{Wardone}:
\begin{align}\label{WID}
W_1:=\sum_{\bm{p}_{\bot_1}}\Big[\bigg(C^{-1}({\bf p})-C^{-1}(a_1{\bf p}_{\bot_1})\bigg)\bigg(G^{(1,1)}_{a_1{\bf p}_{\bot_1};{\bf p}}+G^{(1,0)}_{a_1{\bf p}_{\bot_1}}G^{(0,1)}_{\bf p}\bigg)+\bar{\J}_{a_1{\bf p}_{\bot_1}}G^{(0,1)}_{\bf p}-\J_{\bf p}G^{(1,0)}_{a_1{\bf p}_{\bot_1}}\Big]=0
\end{align}
The partial derivative of this identity with respect to $\J$, in the first time, and with respect to $\bar \J$ in the second time are respectively
\bea\label{fancy1}
\frac{\partial W_i}{\partial \J_{\bf q}}&=&\sum_{\bm{p}_{\bot_1}}\Delta_{C_0}\Big(G^{(2,1)}_{a_1{\bf p}_{\bot_1}{\bf q};\bf p}+G^{(0,1)}_{\bf p}G^{(2,0)}_{a_1{\bf p}_{\bot_1}{\bf q}}+G^{(1,0)}_{a_1{\bf p}_{\bot_1}}G^{(1,1)}_{\bf q;\bf p}\Big)\cr
&+&\sum_{\bm{p}_{\bot_1}}\Big(\bar{\J}_{a_1{\bf p}_{\bot_1}}G^{(1,1)}_{\bf q;\bf p}-G^{(2,0)}_{a_1{\bf p}_{\bot_1}\bf q}\J_{\bf p}-G^{(1,0)}_{a_1{\bf p}_{\bot_1}}\delta_{\bf pq}\Big)=0
\eea
and
\bea\label{fancy2}
\frac{\partial W_i}{\partial \bar{\J}_{\bf q'}}&=&\sum_{\bm{p}_{\bot_1}}\Delta _{C_0}\Big(G^{(1,2)}_{a_1{\bf p}_{\bot_1};\bf p\bf q'}+G^{(0,1)}_{\bf p}G^{(1,1)}_{a_1{\bf p}_{\bot_1};\bf q'}+G^{(1,0)}_{a_1{\bf p}_{\bot_1}}G^{(0,2)}_{\bf p\,\bf q\,'}\Big)\cr
&+&\sum_{\bm{p}_{\bot_1}}\Big(\delta_{a_1{\bf p}_{\bot_1}\bf q'}G^{(0,1)}_{\bf p}+\bar{\J}_{a_1{\bf p}_{\bot_1}}G^{(0,2)}_{\bf p\bf q'}-G^{(1,1)}_{a_1{\bf p}_{\bot_1};\bf q'}\J_{\bf p}\Big)=0,
\eea
where $\Delta _{C_0}=(C^{-1}({\bf p})-C^{-1}(a_1{\bf p}_{\bot_1}))$.
Now multiplying \eqref{fancy1} by $\J_{\bf q' }$ and \eqref{fancy2} by $\bar{\J}_{\bf q}$ and summing  these two expressions leads to the following relation:
\bea\label{malmal}
&&\sum_{\bm{p}_{\bot_1}}\Delta_{C_0}\Big[\J_{\bf q' }\Big(G^{(2,1)}_{a_1{\bf p}_{\bot_1}{\bf q};\bf p}+G^{(0,1)}_{\bf p}G^{(2,0)}_{a_1{\bf p}_{\bot_1}{\bf q}}+G^{(1,0)}_{a_1{\bf p}_{\bot_1}}G^{(1,1)}_{\bf q;\bf p}\Big)\cr
&&+\bar{\J}_{\bf q}\Big(G^{(1,2)}_{a_1{\bf p}_{\bot_1};\bf p\bf q'}+G^{(0,1)}_{\bf p}G^{(1,1)}_{a_1{\bf p}_{\bot_1};\bf q'}+G^{(1,0)}_{a_1{\bf p}_{\bot_1}}G^{(0,2)}_{\bf p\,\bf q\,'}\Big)\Big]\cr
&&+\sum_{\bm{p}_{\bot_1}}\Big(\J_{\bf q' }\bar{\J}_{a_1{\bf p}_{\bot_1}}G^{(1,1)}_{\bf q;\bf p}-\J_{\bf q' }\J_{\bf p}G^{(2,0)}_{a_1{\bf p}_{\bot_1}\bf q}+\bar{\J}_{\bf q}\bar{\J}_{a_1{\bf p}_{\bot_1}}G^{(0,2)}_{\bf p\bf q'}-\bar{\J}_{\bf q}\J_{\bf p}G^{(1,1)}_{a_1{\bf p}_{\bot_1};\bf q'}\Big)\cr
&&+\sum_{\bm{p}_{\bot_1}}\Big(\delta_{a_1{\bf p}_{\bot_1}\bf q'}\bar{\J}_{\bf q}G^{(0,1)}_{\bf p}-\delta_{\bf pq}\J_{\bf q'}G^{(1,0)}_{a_1{\bf p}_{\bot_1}}\Big)=0
\eea
Now multiplying \eqref{fancy1} by $\J_{\bf q' }$ and \eqref{fancy2} by $\bar{\J}_{\bf q}$ and subtract one by the other  leads to the following relation:
\bea\label{malmal}
&&\sum_{\bm{p}_{\bot_1}}\Delta_{C_0}\Big[\J_{\bf q' }\Big(G^{(2,1)}_{a_1{\bf p}_{\bot_1}{\bf q};\bf p}+G^{(0,1)}_{\bf p}G^{(2,0)}_{a_1{\bf p}_{\bot_1}{\bf q}}+G^{(1,0)}_{a_1{\bf p}_{\bot_1}}G^{(1,1)}_{\bf q;\bf p}\Big)\cr
&&-\bar{\J}_{\bf q}\Big(G^{(1,2)}_{a_1{\bf p}_{\bot_1};\bf p\bf q'}+G^{(0,1)}_{\bf p}G^{(1,1)}_{a_1{\bf p}_{\bot_1};\bf q'}+G^{(1,0)}_{a_1{\bf p}_{\bot_1}}G^{(0,2)}_{\bf p\,\bf q\,'}\Big)\Big]\cr
&&+\sum_{\bm{p}_{\bot_1}}\Big(\J_{\bf q' }\bar{\J}_{a_1{\bf p}_{\bot_1}}G^{(1,1)}_{\bf q;\bf p}-\J_{\bf q' }\J_{\bf p}G^{(2,0)}_{a_1{\bf p}_{\bot_1}\bf q}-\bar{\J}_{\bf q}\bar{\J}_{a_1{\bf p}_{\bot_1}}G^{(0,2)}_{\bf p\bf q'}+\bar{\J}_{\bf q}\J_{\bf p}G^{(1,1)}_{a_1{\bf p}_{\bot_1};\bf q'}\Big)\cr
&&-\sum_{\bm{p}_{\bot_1}}\Big(\delta_{\bf pq}\J_{\bf q'}G^{(1,0)}_{a_1{\bf p}_{\bot_1}}+\delta_{a_1{\bf p}_{\bot_1}\bf q'}\bar{\J}_{\bf q}G^{(0,1)}_{\bf p}\Big)=0
\eea
Finally the second order derivative with respect to $\J_{\bf q}$ and $\bar{\J}_{\bf q'}$ is
\bea\label{fancy1fawaaz}
\frac{\partial^2 W_i}{\partial \J_{\bf q}\partial\bar{\J}_{\bf q'}}&=&\sum_{\bm{p}_{\bot_1}}\Delta_{C_0}\Big(G^{(2,2)}_{a_1{\bf p}_{\bot_1}{\bf q};{\bf pq'}}+G^{(0,2)}_{{\bf p}{\bf q'}}G^{(2,0)}_{a_1{\bf p}_{\bot_1}{\bf q}}+G^{(0,1)}_{\bf p}G^{(2,1)}_{a_1{\bf p}_{\bot_1}{\bf q};{\bf q'}}\cr
&+&G^{(1,1)}_{a_1{\bf p}_{\bot_1};{\bf q'}}G^{(1,1)}_{\bf q;p}+G^{(1,0)}_{a_1{\bf p}_{\bot_1}}G^{(1,2)}_{{\bf q};{\bf p q'}}\Big)\cr
&+&\sum_{\bm{p}_{\bot_1}}\Big(\delta_{a_1{\bf p}_{\bot_1}{\bf q'}}G^{(1,1)}_{{\bf q};{\bf p}}+\bar{\J}_{a_1{\bf p}_{\bot_1}}G^{(1,2)}_{{\bf q};{\bf pq' }}-\J_{\bf p}G^{(2,1)}_{a_1{\bf p}_{\bot_1}{\bf q};{\bf q'}}-\delta_{\bf pq}G^{(1,1)}_{a_1{\bf p}_{\bot_1};{\bf q'}}\Big)=0.
\eea
The Ward identity $W_{11}$ (equation \eqref{machiavel}) can be rewritten as:

\bea 
	W_{11} &=& \sum_{{\bm{p}}_{\bot_1}}  \bigg\{\Delta_{C_1}\bigg(G^{(1,1)}_{r_1{\bf p}_{\bot_1};q_1{\bf p}_{\bot_1}}+G^{(0,1)}_{q_1{\bf p}_{\bot_1}}G^{(1,0)}_{r_1{\bf p}_{\bot_1}}\bigg)+\frac{1}{2}\bigg(\bar{\J}_{r_1{\bf p_{\bot_1}}}G^{(0,1)}_{q_1{\bf p}_{\bot_1}}+\J_{q_1{\bf p}_{\bot_1}}G^{(1,0)}_{r_1{\bf p}_{\bot_1}}\bigg)\cr
	&+&\frac{1}{2}\Delta_{C_2}\bigg(G^{(1,0)}_{s{\bf p}_{\bot_1}}G^{(1,1)}_{r_1{\bf p}_{\bot_1};q_1{\bf p}_{\bot_1}}G^{(0,1)}_{s{\bf p}_{\bot_1}}+G^{(2,1)}_{r_1{\bf p}_{\bot_1}s{\bf p}_{\bot_1};q_1{\bf p}_{\bot_1}}G^{(0,1)}_{s{\bf p}_{\bot_1}}+G^{(1,0)}_{s{\bf p}_{\bot_1}}G^{(1,1)}_{r_1{\bf p}_{\bot_1};s{\bf p}_{\bot_1}}G^{(0,1)}_{q_1{\bf p}_{\bot_1}}\cr
	&+&G^{(1,1)}_{s{\bf p}_{\bot_1};q_1{\bf p}_{\bot_1}}G^{(1,1)}_{r_1{\bf p}_{\bot_1};s{\bf p}_{\bot_1}}+
	G^{(1,1)}_{r_1{\bf p}_{\bot_1};q_1{\bf p}_{\bot_1}}G^{(1,1)}_{s{\bf p}_{\bot_1};s{\bf p}_{\bot_1}}+G^{(2,1)}_{r_1{\bf p}_{\bot_1}s{\bf p}_{\bot_1};s{\bf p}_{\bot_1}}G^{(0,1)}_{q_1{\bf p}_{\bot_1}}\cr
	&+&G^{(2,0)}_{s{\bf p}_{\bot_1}r_1{\bf p}_{\bot_1}} G^{(0,1)}_{q_1{\bf p}_{\bot_1}} G^{(0,1)}_{s{\bf p}_{\bot_1}} +G^{(2,0)}_{s{\bf p}_{\bot_1}r_1{\bf p}_{\bot_1}}G^{(0,2)}_{q_1{\bf p}_{\bot_1}s{\bf p}_{\bot_1}}+G^{(1,2)}_{r_1{\bf p}_{\bot_1};s{\bf p}_{\bot_1}q_1{\bf p}_{\bot_1}}G^{(1,0)}_{s{\bf p}_{\bot_1}}\cr
	&+&G^{(1,0)}_{r_1{\bf p}_{\bot_1}}G^{(1,1)}_{s{\bf p}_{\bot_1};q_1{\bf p}_{\bot_1}}G^{(0,1)}_{s{\bf p}_{\bot_1}}+G^{(1,0)}_{r_1{\bf p}_{\bot_1}}G^{(1,1)}_{s{\bf p}_{\bot_1};s{\bf p}_{\bot_1}}G^{(0,1)}_{q_1{\bf p}_{\bot_1}}+G^{(1,0)}_{r_1{\bf p}_{\bot_1}}G^{(1,0)}_{s{\bf p}_{\bot_1}}G^{(0,1)}_{q_1{\bf p}_{\bot_1}}G^{(0,1)}_{s{\bf p}_{\bot_1}}\cr
	&+&G^{(1,0)}_{r_1{\bf p}_{\bot_1}}G^{(1,0)}_{s{\bf p}_{\bot_1}}G^{(0,2)}_{s{\bf p}_{\bot_1}q_1{\bf p}_{\bot_1}}+G^{(1,0)}_{r_1{\bf p}_{\bot_1}}G^{(1,2)}_{s{\bf p}_{\bot_1};q_1{\bf p}_{\bot_1}s{\bf p}_{\bot_1}}+G^{(2,2)}_{r_1{\bf p}_{\bot_1}s{\bf p}_{\bot_1};q_1{\bf p}_{\bot_1}s{\bf p}_{\bot_1}}\bigg)\cr
	&+&\frac{1}{2}\Delta_{C_3}\bigg[-\bar{\J}_{\bf p}\bigg(G^{(1,1)}_{r_1{\bf p}_{\bot_1};q_1{\bf p}_{\bot_1}}G^{(0,1)}_{s{\bf p}_{\bot_1}}+G^{(1,1)}_{r_1{\bf p}_{\bot_1};s{\bf p}_{\bot_1}}G^{(0,1)}_{q_1{\bf p}_{\bot_1}}+G^{(1,0)}_{r_1{\bf p}_{\bot_1}}G^{(0,1)}_{q_1{\bf p}_{\bot_1}}G^{(0,1)}_{s{\bf p}_{\bot_1}}\cr
	&+&G^{(1,0)}_{r_1{\bf p}_{\bot_1}}G^{(0,2)}_{s{\bf p}_{\bot_1}q_1{\bf p}_{\bot_1}}+G^{(1,2)}_{r_1{\bf p}_{\bot_1};s{\bf p}_{\bot_1}q_1{\bf p}_{\bot_1}}\bigg)+\J_{\bf p}\bigg(G^{(2,0)}_{r_1{\bf p}_{\bot_1}s{\bf p}_{\bot_1}}G^{(0,1)}_{s{\bf p}_{\bot_1}}+G^{(1,1)}_{s{\bf p}_{\bot_1};s{\bf p}_{\bot_1}}G^{(1,0)}_{r_1{\bf p}_{\bot_1}}\cr
	&+&G^{(1,0)}_{s{\bf p}_{\bot_1}}G^{(1,0)}_{r_1{\bf p}_{\bot_1}}G^{(0,1)}_{s{\bf p}_{\bot_1}}+G^{(1,1)}_{r_1{\bf p}_{\bot_1};s{\bf p}_{\bot_1}}G^{(1,0)}_{s{\bf p}_{\bot_1}}+G^{(2,1)}_{r_1{\bf p}_{\bot_1}s{\bf p}_{\bot_1};s{\bf p}_{\bot_1}}\bigg)\bigg]\cr
	&+&\frac{1}{2}\Delta_{C_4}\bigg[-\bar{\J}_{\bf p}\bigg(G^{(1,1)}_{s{\bf p}_{\bot_1};s{\bf p}_{\bot_1}}G^{(0,1)}_{q_1{\bf p}_{\bot_1}}+G^{(1,1)}_{s{\bf p}_{\bot_1};q_1{\bf p}_{\bot_1}}G^{(0,1)}_{s{\bf p}_{\bot_1}}+G^{(1,0)}_{s{\bf p}_{\bot_1}}G^{(0,1)}_{s{\bf p}_{\bot_1}}G^{(0,1)}_{q_1{\bf p}_{\bot_1}}\cr
	&+&G^{(1,0)}_{s{\bf p}_{\bot_1}}G^{(0,2)}_{s{\bf p}_{\bot_1}q_1{\bf p}_{\bot_1}}+G^{(1,2)}_{s{\bf p}_{\bot_1};s{\bf p}_{\bot_1}q_1{\bf p}_{\bot_1}}\bigg)+\J_{\bf p}\bigg(G^{(2,0)}_{s{\bf p}_{\bot_1}r_1{\bf p}_{\bot_1}}G^{(0,1)}_{q_1{\bf p}_{\bot_1}}+G^{(1,0)}_{s{\bf p}_{\bot_1}}G^{(1,1)}_{r_1{\bf p}_{\bot_1};q_1{\bf p}_{\bot_1}}\cr
	&+&G^{(1,0)}_{r_1{\bf p}_{\bot_1}}G^{(1,0)}_{s{\bf p}_{\bot_1}}G^{(0,1)}_{q_1{\bf p}_{\bot_1}}+G^{1,0}_{r_1{\bf p}_{\bot_1}}G^{(1,1)}_{s{\bf p}_{\bot_1};q_1{\bf p}_{\bot_1}}+G^{(2,1)}_{s{\bf p}_{\bot_1}r_1{\bf p}_{\bot_1};q_1{\bf p}_{\bot_1}}\bigg)\Bigg]\cr
	&+&\frac{1}{2}\Bigg[\bar{\J}_{\bf p}\bar{\J}_{\bf p}\bigg(G^{(0,2)}_{s{\bf p}_{\bot_1};q_1{\bf p}_{\bot_1}}+G^{(0,1)}_{s{\bf p}_{\bot_1}}G^{(0,1)}_{q_1{\bf p}_{\bot_1}}\bigg)-\bar{\J}_{\bf p}\J_{\bf p}\bigg(G^{(1,1)}_{s{\bf p}_{\bot_1};{s{\bf p}_{\bot_1}}}+G^{(1,0)}_{s{\bf p}_{\bot_1}}G^{(0,1)}_{s{\bf p}_{\bot_1}}\bigg)\cr
	&-&\J_{\bf p}\bar{\J}_{\bf p}\bigg(G^{(1,1)}_{r_1{\bf p}_{\bot_1};{q_1{\bf p}_{\bot_1}}}+G^{(1,0)}_{r_1{\bf p}_{\bot_1}}G^{(0,1)}_{q_1{\bf p}}\bigg)+\J_{\bf p}\J_{\bf p}\bigg(G^{(2,0)}_{r_1{\bf p}_{\bot_1}s{\bf p}_{\bot_1}}+G^{(1,0)}_{r_1{\bf p}_{\bot_1}}G^{(1,0)}_{s{\bf p}_{\bot_1}}\bigg)\Bigg]\bigg\}.
	\eea 
Taking into account the identity \eqref{machiavel}, we note that the sum $\sum_{{\bf p}_{\bot_1}}\mathcal{D}$, where $\mathcal{D}$ is given by 
 \bea\label{doudou}
 \mathcal{D}&=&\frac{1}{2}\bigg(\bar{\J}_{r_1{\bf p_{\bot_1}}}G^{(0,1)}_{q_1{\bf p}_{\bot_1}}+\J_{q_1{\bf p}_{\bot_1}}G^{(1,0)}_{r_1{\bf p}_{\bot_1}}\bigg)\cr
 &+&\frac{1}{2}\Bigg[\bar{\J}_{\bf p}\bar{\J}_{\bf p}\bigg(G^{(0,2)}_{s{\bf p}_{\bot_1};q_1{\bf p}_{\bot_1}}+G^{(0,1)}_{s{\bf p}_{\bot_1}}G^{(0,1)}_{q_1{\bf p}_{\bot_1}}\bigg)-\bar{\J}_{\bf p}\J_{\bf p}\bigg(G^{(1,1)}_{s{\bf p}_{\bot_1};{s{\bf p}_{\bot_1}}}+G^{(1,0)}_{s{\bf p}_{\bot_1}}G^{(0,1)}_{s{\bf p}_{\bot_1}}\bigg)\cr
	&-&\J_{\bf p}\bar{\J}_{\bf p}\bigg(G^{(1,1)}_{r_1{\bf p}_{\bot_1};{q_1{\bf p}_{\bot_1}}}+G^{(1,0)}_{r_1{\bf p}_{\bot_1}}G^{(0,1)}_{q_1{\bf p}}\bigg)+\J_{\bf p}\J_{\bf p}\bigg(G^{(2,0)}_{r_1{\bf p}_{\bot_1}s{\bf p}_{\bot_2}}+G^{(1,0)}_{r_1{\bf p}_{\bot_1}}G^{(1,0)}_{s{\bf p}_{\bot_1}}\bigg)\Bigg],
 \eea
 is considered as the second-order Ward identity in the case where $\Delta_{C_i}=0$ and is shown to vanish in this situation. To be more precise, taking \eqref{malmal} and setting $\Delta_{C_0}=0$ we have:
 \bea
&&\sum_{\bm{p}_{\bot_1}}\Big(\J_{\bf q' }\bar{\J}_{a_1{\bf p}_{\bot_1}}G^{(1,1)}_{\bf q;\bf p}-\J_{\bf q' }\J_{\bf p}G^{(2,0)}_{a_1{\bf p}_{\bot_1}\bf q}-\bar{\J}_{\bf q}\bar{\J}_{a_1{\bf p}_{\bot_1}}G^{(0,2)}_{\bf p\bf q'}+\bar{\J}_{\bf q}\J_{\bf p}G^{(1,1)}_{a_1{\bf p}_{\bot_1};\bf q'}\Big)\cr
&&-\sum_{\bm{p}_{\bot_1}}\Big(\delta_{\bf pq}\J_{\bf q'}G^{(1,0)}_{a_1{\bf p}_{\bot_1}}+\delta_{a_1{\bf p}_{\bot_1}\bf q'}\bar{\J}_{\bf q}G^{(0,1)}_{\bf p}\Big)=0.
 \eea
Setting ${\bf q}'={\bf q}={\bf p}$, $a_1=p_1=r_1=s=q_1$ in the above relation, the sum of $\mathcal{D}$ given in \eqref{doudou} is reduced to
 \bea
 \sum_{\bf p_{\bot_1}}\mathcal{D}&=&\frac{1}{2}\sum_{\bf p_{\bot_1}}\Big(\bar{\J}_{\bf p}\bar{\J}_{\bf p}G^{(0,1)}_{s{\bf p}_{\bot_1}}G^{(0,1)}_{q_1{\bf p}_{\bot_1}}-\bar{\J}_{\bf p}\J_{\bf p}G^{(1,0)}_{s{\bf p}_{\bot_1}}G^{(0,1)}_{s{\bf p}_{\bot_1}}\cr
	&-&\J_{\bf p}\bar{\J}_{\bf p}G^{(1,0)}_{r_1{\bf p}_{\bot_1}}G^{(0,1)}_{q_1{\bf p}}+\J_{\bf p}\J_{\bf p}G^{(1,0)}_{r_1{\bf p}_{\bot_1}}G^{(1,0)}_{s{\bf p}_{\bot_1}}\Big)
 \eea
 which is nothing but  the combination of the first order Ward identity for $\Delta_{C_0}=0$ and then is identically vanished,
 see \cite{Lahoche:2021nba} for more detail. Now,
using the first order Ward identity \eqref{WID} we can deduce that 
\bea\label{xxl1}
\sum_{{\bf p}_{\bot_1}\in[-\Lambda,\Lambda]^{d-1}} \Delta_{C_1}\cI_1=\frac{2s^2-q_1^2-r_1^2}{2(q_1^2-r_1^2)}\sum_{{\bf p}_{\bot_1}\in[-\Lambda,\Lambda]^{d-1}}\Big(\J_{q_1{\bf p}_{\bot_1}}G^{(1,0)}_{r_1{\bf p}_{\bot_1}}-\bar{\J}_{r_1{\bf p}_{\bot_1}}G^{(0,1)}_{q_1{\bf p}_{\bot_1}}\Big),
\eea
where we modified the regularization procedure on the propagator by directly regulating the momentum component with the cutoff $\Lambda$.
Also using \eqref{WID}, \eqref{fancy1}, \eqref{fancy2} and \eqref{malmal} we have
\bea\label{xxl2}
\sum_{{\bf p}_{\bot_1}\in[-\Lambda,\Lambda]^{d-1}} \Delta_{C_2}\cI_2&=&-\frac{(s^2-r_1^2)(q_1-s^2)}{2(q^2_1-r^2_1)}\sum_{\bm{p}_{\bot_1}\in[-\Lambda,\Lambda]^{d-1}}\Big(\delta_{r_1s}G^{(1,1)}_{s{\bf p}_{\bot_1};q_1{\bf p}_{\bot_1}}+\bar{\J}_{r_1{\bf p}_{\bot_1}}G^{(1,2)}_{s{\bf p}_{\bot_1};q_1{\bf p}_{\bot_1}s{\bf p}_{\bot_1}}\cr
&-&\J_{q_1{\bf p}_{\bot_1}}G^{(2,1)}_{r_1{\bf p}_{\bot_1}s{\bf p}_{\bot_1};s{\bf p}_{\bot_1}}-\delta_{\bf sq_1}G^{(1,1)}_{r_1{\bf p}_{\bot_1};s{\bf p}_{\bot_1}}\Big)\cr
&+&\sum_{{\bf p}_{\bot_1}}\frac{1}{2}\Delta_{C_2}\bigg[\Big(G^{(0,1)}_{s{\bf p}_{\bot_1}}G^{(1,0)}_{s{\bf p}_{\bot_1}}+G^{(1,1)}_{s{\bf p}_{\bot_1};s{\bf p}_{\bot_1}}\Big)\Big(G^{(1,1)}_{r_1{\bf p}_{\bot_1};q_1{\bf p}_{\bot_1}}+G^{(1,0)}_{r_1{\bf p}_{\bot_1}}G^{(0,1)}_{q_1{\bf p}_{\bot_1}}\Big)\cr
&+&G^{(0,1)}_{s{\bf p}_{\bot_1}}\Big(G^{(2,1)}_{r_1{\bf p}_{\bot_1}s{\bf p}_{\bot_1};q_1{\bf p}_{\bot_1}}+G^{(2,0)}_{s{\bf p}_{\bot_1}r_1{\bf p}_{\bot_1}} G^{(0,1)}_{q_1{\bf p}_{\bot_1}}  +G^{(1,0)}_{r_1{\bf p}_{\bot_1}}G^{(1,1)}_{s{\bf p}_{\bot_1};q_1{\bf p}_{\bot_1}}\Big)\cr
	&+&G^{(1,0)}_{s{\bf p}_{\bot_1}}\Big(G^{(1,2)}_{r_1{\bf p}_{\bot_1};s{\bf p}_{\bot_1}q_1{\bf p}_{\bot_1}}+G^{(1,1)}_{r_1{\bf p}_{\bot_1};s{\bf p}_{\bot_1}}G^{(0,1)}_{q_1{\bf p}_{\bot_1}}+G^{(1,0)}_{r_1{\bf p}_{\bot_1}}G^{(0,2)}_{s{\bf p}_{\bot_1}q_1{\bf p}_{\bot_1}}\Big)\bigg],
\eea
\bea
\label{xxl3}
\sum_{{\bf p}_{\bot_1}\in[-\Lambda,\Lambda]^{d-1}} \Delta_{C_3}\cI_3&=&\frac{1}{2}\sum_{{\bf p}_{\bot_1}\in[-\Lambda,\Lambda]^{d-1}} \Delta_{C_3}\bigg[-\bar{\J}_{\bf p}\bigg(G^{(1,1)}_{r_1{\bf p}_{\bot_1};s{\bf p}_{\bot_1}}+G^{(1,0)}_{r_1{\bf p}_{\bot_1}}G^{(0,1)}_{s{\bf p}_{\bot_1}}\bigg)G^{(0,1)}_{q_1{\bf p}_{\bot_1}}\cr
&+&\J_{\bf p}\bigg(G^{(1,1)}_{r_1{\bf p}_{\bot_1},s{\bf p}_{\bot_1}}+G^{(1,0)}_{r_1{\bf p}_{\bot_1}}G^{(0,1)}_{s{\bf p}_{\bot_1}}\bigg)G^{(1,0)}_{s{\bf p}_{\bot_1}}\bigg]\cr
&+&\frac{1}{2}\sum_{\bm{p}_{\bot_1}}\Big(\J_{q_1{\bf p}_{\bot_1} }\bar{\J}_{r_1{\bf p}_{\bot_1}}G^{(1,1)}_{s{\bf p}_{\bot_1};s{\bf p}_{\bot_1}}-\J_{q_1{\bf p}_{\bot_1} }\J_{s{\bf p}_{\bot_1}}G^{(2,0)}_{r_1{\bf p}_{\bot_1}s{\bf p}_{\bot_1}}\cr
&-&\bar{\J}_{s{\bf p}_{\bot_1}}\bar{\J}_{r_1{\bf p}_{\bot_1}}G^{(0,2)}_{s{\bf p}_{\bot_1}q_1{\bf p}_{\bot_1}}+\bar{\J}_{s{\bf p}_{\bot_1}}\J_{s{\bf p}_{\bot_1}}G^{(1,1)}_{r_1{\bf p}_{\bot_1};q_1{\bf p}_{\bot_1}}\Big)\cr
&-&\frac{1}{2}\sum_{\bm{p}_{\bot_1}}\Big(\J_{q_1{\bf p}_{\bot_1}}G^{(1,0)}_{r_1{\bf p}_{\bot_1}}+\delta_{r_1q_1}\bar{\J}_{s{\bf p}_{\bot_1}}G^{(0,1)}_{s{\bf p}_{\bot_1}}\Big),
\eea
\bea\label{xxl4}
\sum_{{\bf p}_{\bot_1}\in[-\Lambda,\Lambda]^{d-1}} \Delta_{C_4}\cI_4&=&\frac{1}{2}\sum_{{\bf p}_{\bot_1}\in[-\Lambda,\Lambda]^{d-1}} \Delta_{C_4}\bigg[-\bar{\J}_{\bf p}\bigg(G^{(1,1)}_{s{\bf p}_{\bot_1};q_1{\bf p}_{\bot_1}}+G^{(1,0)}_{s{\bf p}_{\bot_1}}G^{(0,1)}_{q_1{\bf p}_{\bot_1}}\bigg)G^{(0,1)}_{s{\bf p}_{\bot_1}}\cr
&+&\J_{\bf p}\bigg(G^{(1,1)}_{s{\bf p}_{\bot_1},q_1{\bf p}_{\bot_1}}+G^{(1,0)}_{s{\bf p}_{\bot_1}}G^{(0,1)}_{q_1{\bf p}_{\bot_1}}\bigg)G^{(1,0)}_{r_1{\bf p}_{\bot_1}}\bigg]\cr
&+&\frac{1}{2}\sum_{\bm{p}_{\bot_1}}\Big(\J_{s{\bf p}_{\bot_1}}\bar{\J}_{s{\bf p}_{\bot_1}}G^{(1,1)}_{r_1{\bf p}_{\bot_1};q_1{\bf p}_{\bot_1}}-\J_{s{\bf p}_{\bot_1} }\J_{q_1{\bf p}_{\bot_1}}G^{(2,0)}_{s{\bf p}_{\bot_1}r_1{\bf p}_{\bot_1}}\cr
&-&\bar{\J}_{r_1{\bf p}_{\bot_1}}\bar{\J}_{s{\bf p}_{\bot_1}}G^{(0,2)}_{q_1{\bf p}_{\bot_1}s{\bf p}_{\bot_1}}+\bar{\J}_{r_1{\bf p}_{\bot_1}}\J_{q_1{\bf p}_{\bot_1}}G^{(1,1)}_{s{\bf p}_{\bot_1};s{\bf p}_{\bot_1}}\Big)\cr
&-&\frac{1}{2}\sum_{\bm{p}_{\bot_1}}\Big(\delta_{r_1q_1}\J_{s{\bf p}_{\bot_1}}G^{(1,0)}_{s{\bf p}_{\bot_1}}+\bar{\J}_{r_1{\bf p}_{\bot_1}}G^{(0,1)}_{q_1{\bf p}_{\bot_1}}\Big).
\eea
Finally the second order Ward identity should be of the form 
$
W_{11}=f(W_1)+ \mathcal{T}_{11}
$
where $f(W_1)$ is vanished function as soon as $W_1=0$. $\mathcal{T}_{11}$ is a sum of the expression \eqref{xxl1}, \eqref{xxl2}, \eqref{xxl3}, \eqref{xxl4}. Now we come to the following claim:
\begin{equation}\label{tomjerry}
\{ \Delta_{C_i}=0\,\, \forall i=1,2,3,4\} \Longleftrightarrow \mathcal{T}_{11}=0.
\end{equation}
Let us provide the proof of relation \eqref{tomjerry}. First assuming that $\Delta_{ C_i}=0$, then identity $W_{11}=\mathcal{D}=0$ which is identical to the case where we impose the closure constraint to our model. For more detail see \cite{Lahoche:2021nba}. Then assuming that $\mathcal{T}_{11}=0$ and $\Delta_{C_i}\neq 0$. Due to the form of $\Delta_{C_i}$ all the expressions \eqref{doudou},   \eqref{xxl1},  \eqref{xxl2}, \eqref{xxl3} and \eqref{xxl4} are independent, and  we can remark that this statement implies that 
\bea
G^{(1,1)}_{\bf p;q}+G^{(1,0)}_{\bf p}G^{(0,1)}_{\bf q}=0,
\eea
\bea
G^{(2,1)}_{r_1{\bf p}_{\bot_1}s{\bf p}_{\bot_1};q_1{\bf p}_{\bot_1}}+G^{(2,0)}_{s{\bf p}_{\bot_1}r_1{\bf p}_{\bot_1}} G^{(0,1)}_{q_1{\bf p}_{\bot_1}}  +G^{(1,0)}_{r_1{\bf p}_{\bot_1}}G^{(1,1)}_{s{\bf p}_{\bot_1};q_1{\bf p}_{\bot_1}}=0,
\eea
and
\bea
G^{(1,2)}_{r_1{\bf p}_{\bot_1};s{\bf p}_{\bot_1}q_1{\bf p}_{\bot_1}}+G^{(1,1)}_{r_1{\bf p}_{\bot_1};s{\bf p}_{\bot_1}}G^{(0,1)}_{q_1{\bf p}_{\bot_1}}+G^{(1,0)}_{r_1{\bf p}_{\bot_1}}G^{(0,2)}_{s{\bf p}_{\bot_1}q_1{\bf p}_{\bot_1}}=0,
\eea
which violates the first order Ward identity (see \eqref{WID}, \eqref{fancy1}, \eqref{fancy2}) and therefore $\mathcal{T}_{11}=0$ implies that $\Delta_{C_i}=0$. Finally the claim \eqref{tomjerry} is well satisfied and the same conclusion holds for $W_{12}$ and $\tilde{W}_{11}$.

\section{Concluding remarks}\label{thirdsection}
We provide here some comments and conclude our work. First, let us remark that the first order Ward identity \eqref{WID} expressed in terms of the observables $G$ by taking $\J=\bar{\J}=0$, takes the simple form
\bea\label{mianmian}
W_1=(p_1^2-a_1^2)\sum_{\bm{p}_{\bot_1}\in[-\Lambda,\Lambda]^{d-1}}\Big(G^{(1,1)}_{a_1{\bf p}_{\bot_1};{\bf p}}+G^{(1,0)}_{a_1{\bf p}_{\bot_1}}G^{(0,1)}_{\bf p}\Big)=0.
\eea
In the symmetric phase where the odd correlation functions vanish, the identity \eqref{mianmian} becomes,
$
(p_1^2-a_1^2)\sum_{\bm{p}_{\bot_1}\in[-\Lambda,\Lambda]^{d-1}}G^{(1,1)}_{a_1{\bf p}_{\bot_1};\,p_1{\bf p}_{\bot_1}}=0,
$
which have the only solution $G^{(1,1)}_{a_1{\bf p}_{\bot_1};\,p_1{\bf p}_{\bot_1}}\propto \delta_{a_1p_1}$ and is in agreement with the symmetric phase assumption $G^{(1,1)}_{\bf p;q}\propto \delta_{\bf pq}$.
\medskip

Second and as explained above, despite the fact that we expect that all the relations between the correlation functions coming from the first-order Ward identity \eqref{WID} have to be redundant at second and higher orders. This is due to the group structure of the transformation. It seems that this is not the case after all. We have shown explicitly that this assumption fails in this paper: the group argument is undermined, and the second-order Ward identities represent new symmetry relations to be considered. In our previous work \cite{Lahoche:2021nba} we have shown that models with closure constraint do not present such an anomaly. At this stage, it should be noticed that this anomaly does not seem to be intrinsically related to tensor, but concern also vector or matrix field theories in zero dimension with free propagator behaving like $1/p^2$. Indeed, according to our analysis, it seems that the origin of this phenomenon is to be found in the kinetic term and not in the non-locality of the interactions, which are precisely constructed to remain invariant under global unitary transformations. \medskip

\paragraph{Acknowledgement} V.L. would like to thank "the little crab" for his inspiration. 


\printbibliography[heading=bibintoc]
\end{document}